\documentclass[showpacs,amsmath,amsfonts,amssymb,aps,superscriptaddress]{revtex4}

\usepackage{tabularx}
\usepackage{comment}
\usepackage{enumitem}
\usepackage{graphicx}
\usepackage{float}
\usepackage{dcolumn}
\usepackage{bm}
\usepackage{mathrsfs}
\usepackage{amsmath}

\usepackage{amsfonts}
\usepackage{dsfont}

\begin{document}

\title{Quasinormal Modes in Noncommutative Schwarzschild Black Holes: A Spectral Analysis}
\author{Davide Batic}
\email{davide.batic@ku.ac.ae}
\affiliation{
Mathematics Department, Khalifa University of Science and Technology, PO Box 127788, Abu Dhabi, United Arab Emirates}
\author{Denys Dutykh}
\email{denys.dutykh@ku.ac.ae}
\affiliation{
Mathematics Department, Khalifa University of Science and Technology, PO Box 127788, Abu Dhabi, United Arab Emirates}
\affiliation{Causal Dynamics Pty Ltd, Perth, Australia}

\date{\today}

\begin{abstract}
We present a comprehensive analysis of quasinormal modes (QNMs) for noncommutative geometry-inspired Schwarzschild black holes, encompassing both non-extreme and extreme cases. By employing a spectral method, we calculate the QNMs in the context of scalar, electromagnetic, and gravitational perturbations. Our findings not only challenge previous claims in the literature regarding the instability of these black holes but also reveal remarkable stability for both non-extreme and extreme Schwarzschild black holes under various perturbations.
\end{abstract}
\maketitle

\section{Introduction}

What exactly is a quasinormal mode (QNM)? In essence, we can think of it as the real-world counterpart to the idealized normal mode. To illustrate this concept, let us consider the example of a gently perturbed wine glass. When we tap the glass lightly, it produces a sound that gradually fades away as energy dissipates into the surroundings. If this were a normal mode, the ringing would persist indefinitely. In our current study, we venture into a realm of objects far more intriguing than an ordinary wine glass, namely, black holes. When we delve into more complex scenarios, accounting for possible dissipative phenomena, it becomes natural to expect that an expansion in normal modes is no longer feasible. In such cases, it will be quite natural for the solutions to exhibit decay over time instead of oscillating indefinitely. For this reason, the frequencies will be represented by complex numbers. Let us consider the possibility that the nature of the equilibrium point of a physical system around which we perform an expansion is not known a priori. The QNMs we determine may be characterized by frequencies whose imaginary part does not describe the damping of solutions around the aforementioned point but instead signifies an exponential growth over time. It is easy to deduce that, in this case, the equilibrium point is, in fact, unstable. The significance of this concept is profound: often, the expansion in QNMs is employed to ascertain the stability of highly complex systems, and it was precisely this idea that prompted Regge and Wheeler in their seminal paper \cite{Regge1957PR}  to initiate the study of black holes within the perturbative framework.

Over the years, various techniques have been developed to study the QNMs, each with its own peculiarities. One of the most important methods for determining them is the WKB (Wentzel--Kramers--Brillouin) method \cite{Hall2013}. This particular approach extensively exploits the analogy between the physical phenomenon we are studying here and the study of energy eigenvalues for wave functions using the Schr\"{o}dinger equation in Quantum Mechanics. Another technique useful for the determination of QNMs makes use of group theory. The proposed technique, known as the inverted potential method \cite{Ferrari1984PRD}, is specifically applicable only in the case of a Schwarzschild black hole, not a Kerr black hole. It is based on the similarity between the effective and reversed P\"{o}schl--Teller potentials. Furthermore, the continued fractions method analysis was first introduced by \cite{Leaver1985PRSLA, Leaver1986PRD} while the Laplace transform method was exploited in \cite{Nollert1992PRD}. 
In addition to the aforementioned techniques, it is worth mentioning that the Asymptotic Iterative Method (AIM) has also been applied to determine the QNMs for Schwarzschild, Reissner-Nordström, and Kerr black holes in the case of spin $0$, $1/2$, and $2$ perturbations \cite{Ciftci2003JPA,Cho2010CQG,Cho2012AMP,Batic2018PRD,Panosso2019PRD,Batic2019PRDa,Ciftci2005PLA}. Recently, in a study by \cite{Mamani2022EPJC}, QNMs of a Schwarzschild black hole were revisited using an innovative approach called the spectral method. Spectral methods, though less common in the general relativity (GR) quasinormal mode (QNM) community, offer substantial advantages both analytically and numerically. These methods harness spectral expansions derived from constructive approximation theory. The coefficients within these expansions are computed numerically, resulting in an approximate analytical representation of the solution, effectively blending the strengths of both analytical and numerical approaches.

In contrast to widely used methods such as WKB, inverse potential, and AIM methods, spectral techniques do not rely on local approximations of the potential near its maximum nor require constraints on the angular momentum parameter. The WKB technique also struggles with potentials whose second derivative at the maximum is zero, thus leading to divergent behaviours in formulae like (4.1) in \cite{IYER1987PRD}. Finally, for a thorough analysis of the validity of the WKB approximation in calculating the asymptotic quasinormal modes of black holes we refer to \cite{Daghigh2012PRD}. Asymptotic iteration methods, while useful, are highly sensitive to the choice of initial values in the parameter space. An incorrect choice during the initialization process can affect the stability of the method and the number of iterations required to compute eigenvalues or quasinormal modes. A recent attempt to find the optimal selection can be found in \cite{Batic2023MMAS}.

On the other hand, spectral methods utilize the full analytical form of the potential without any need for local approximations or heuristic modifications. This comprehensive approach not only provides high accuracy but also ensures exponentially fast convergence for the regular part of the solution of a radial equation once its singular behaviour has been completely determined. This feature is particularly advantageous in the search of QNMs of astrophysical objects. For the aforementioned reasons, in the current paper, we will employ an improved version of the spectral technique to explore the QNMs of the Schwarzschild black hole inspired by noncommutative geometry in the case of a massless particle of spin $s = 0$ (scalar perturbation), $s = 1$ (electromagnetic perturbation), and $s = 2$ (vector-type gravitational perturbation) \cite{Nicolini2006PLB}.

While a substantial body of literature exists dedicated to the investigation of QNMs in the context of the classic Schwarzschild spacetime \cite{Bachelot1993AIHPPT, Gundlach1994PRD, Andersson1995PRD, Andersson1997PRD,Konoplya2003ATMP,Motl2003ATMP, Natario2004ATMP, Chakrabarti2006IJMPA, Batic2018PRD}, the situation is considerably less comprehensive for the noncommutative Schwarzschild manifold. In the latter case, the literature contains only partial and occasionally contradictory findings, which we will briefly address in the following discussion.  For example, \cite{Giri2007IJMPA} analyzed the asymptotic QNMs of the nonextreme black hole inspired by noncommutative geometry under the assumption that the black hole mass significantly exceeds its extreme mass. As previously highlighted by \cite{Batic2019EPJC}, in this regime, the QNMs already align with those of its classical counterpart. Hence, it comes as no surprise that \cite{Giri2007IJMPA} concluded that the asymptotic QNMs remain proportional to $\ln{3}$, thus mirroring the corresponding behaviour in the case of the classic Schwarzschild solution. Recently, \cite{Liang2018CPLa, Liang2018CPLb} conducted a study on QNMs of massless scalar field perturbations in the nonextreme regime utilizing the third-order WKB approximation. The author specifically focused on cases with $\ell \in \{1,2,3\}$. In this work, the quasinormal frequencies have been computed for values of the noncommutative parameter $\theta$ across the interval $[0.01, 0.2758]$. If we recall that $\sqrt{\theta}$ represents a quantum of length, it is reasonable to assume that $\sqrt{\theta}$ is on the order of the Planck length, $L_P$. Consequently, the interval chosen in \cite{Liang2018CPLa, Liang2018CPLb} corresponds to a selection of black hole masses within the range $3.6 M_P < M < 19 M_P$, where $M_P$ represents the Planck mass. In this specific range and employing the third-order WKB approximation, \cite{Liang2018CPLa, Liang2018CPLb} observed stable results for the QNMs. It is noteworthy that, in contrast, \cite{Batic2019EPJC} identified instabilities using WKB approximations from third sixth order not only in the narrower range of $1.9 M_P < M < 2.3897 M_P$ but also in the extreme case $\mu=1.9 M_P$. \cite{Batic2019EPJC} emphasized that extending the WKB calculations to the sixth order reveals a lack of convergence precisely when the presumed instabilities occur. To gain insight into the nature of these instabilities, \cite{Batic2019EPJC}  resorted to alternative methods. The inverted potential and Asymptotic Iterative Methods (AIM) failed to detect any instability. However, it is worth noting that both of these methods entail certain approximations, and the AIM method, in particular, is known to encounter convergence issues. For a more in-depth discussion of this aspect, we refer to \cite{Batic2023MMAS}. Last but certainly not least, \cite{Batic2010PLB} conducted an examination of the stability within the interior of the noncommutative Schwarzschild black hole. This analysis involved the study of massless scalar field propagation between the two horizons. Their findings demonstrated that spacetime fuzziness, influenced by higher momenta of the field, effectively mitigates the classical exponential blue shift divergence. This mitigation suppresses the emergence of infinite energy density in regions proximate to the Cauchy horizon. Conversely, \cite{Brown2011PLB} investigated the surface gravity of both inner and outer horizons and posited that the Cauchy horizon might become unstable when subjected to perturbations from infalling matter. In light of this ongoing discussion, there is a significant interest in investigating the (in)stabilities of nonextreme and extreme noncommutative geometry-inspired black holes by analyzing their respective QNMs. In the present study, we tackle this challenge by employing the spectral method to analyze quasinormal frequencies associated with scalar, electromagnetic, and vector-type gravitational perturbations. Our results unequivocally demonstrate the absence of instabilities in both the extreme and nonextreme cases. Notably, in the extreme case, where the event horizon essentially serves as a de facto Cauchy horizon, our findings align with the conclusions derived in \cite{Batic2010PLB}, further bolstering their validity.

The paper is organised as follows. In Section II, we introduce a suitable rescaling of the noncommutative Schwarzschild manifold and derive the effective potential for a massless scalar field in the aforementioned geometry. Moreover, we also establish the corresponding QNMs boundary conditions of a massless scalar field in the presence of nonextreme and extreme noncommutative Schwarzschild black holes. Section III describes the numerical method used in our study, providing the basis for our computational approach. In Section IV, we validate our numerical method by applying it to large-mass noncommutative Schwarzschild black holes. This choice is strategic as, in this high-mass regime, these black holes converge towards the characteristics of classical Schwarzschild black holes, providing a robust benchmark for validation. Here, we also delve into the numerical findings for both nonextreme and extreme noncommutative black holes. Finally, Section V offers conclusions and an outlook, reflecting on the implications of our findings and suggesting directions for future research.

\section{Equations of motion}

\begin{figure}
\includegraphics[scale=0.35]{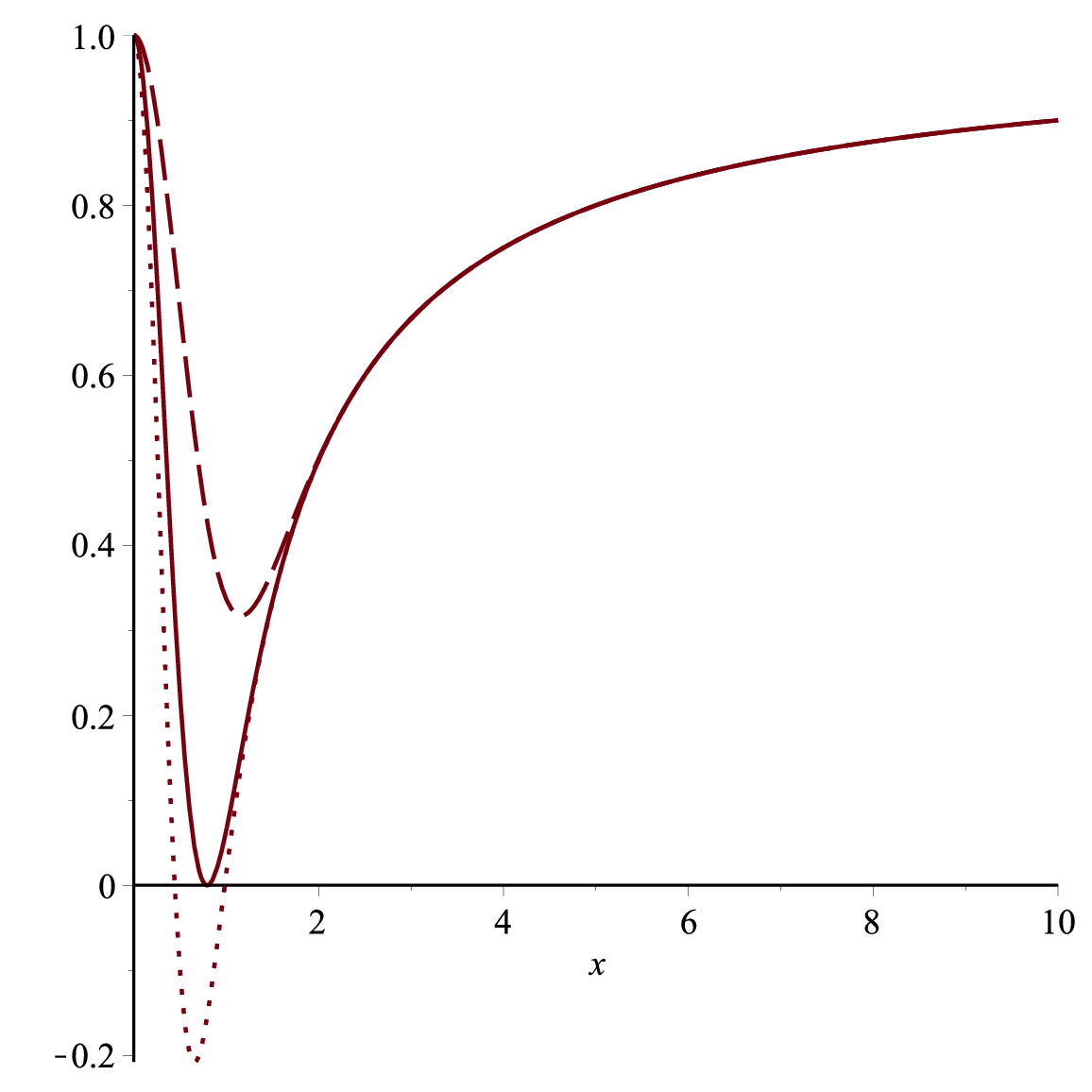}
\caption{\label{fig0}
Plot of the metric coefficient $g_{00} = f(x)$ defined in (\ref{f}). An extreme black hole occurs when $\mu_e= 1.904119076\ldots$ (solid line) and the horizon is located at $x_e= 0.7936575898\ldots$ A nonextreme black hole with two horizons is present for $\mu > \mu_e$ (see the dotted line corresponding to the choice $\mu = 2.3$). A naked gravitational droplet occurs when $0 < \mu < \mu_e$ (see the dashed line for $\mu = 1.3$).}
\end{figure}

We consider a massless scalar field $\psi$ immersed in the noncommutative geometry inspired Schwarzschild background whose line element in units where $c = G_N = 1$ reads \cite{Nicolini2006PLB}
\begin{equation}\label{metric}
 ds^2=-f(r)dt^2+\frac{dr^2}{f(r)} + r^2d\vartheta^2 + r^2\sin^2{\vartheta}d\varphi^2, \quad \vartheta\in[0,\pi], \quad \varphi\in[0,2\pi[, \quad\quad f(r)=1-\frac{2m(r)}{r} 
\end{equation}
with the mass function
\begin{equation}
  m(r) = \frac{2M}{\sqrt{\pi}}\gamma\left(\frac{3}{2},\frac{r^2}{4\theta}\right), \qquad 
  \gamma\left(\frac{3}{2},\frac{r^2}{4\theta}\right) = \int_0^{r^2/4\theta}dt\sqrt{t} e^{-t},
\end{equation}
where $M$ is the total mass of a gravitational object, $\theta$ is a parameter encoding noncommutativity and having the dimension of a length squared. At the same time, $\gamma(\cdot,\cdot)$ is the incomplete lower gamma function. Considering the relationship between the upper and lower incomplete gamma functions as expressed by the formula \cite{Abramowitz1972}
\begin{equation}\label{conversion}
  \gamma\left(\frac{3}{2},\frac{r^2}{4\theta}\right)+\Gamma\left(\frac{3}{2},\frac{r^2}{4\theta}\right)=\frac{\sqrt{\pi}}{2},
\end{equation}
we can represent the $g_{00}$ metric coefficient in (\ref{metric}) as a combination of the typical Schwarzschild term augmented by a noncommutativity-induced perturbation, namely
\begin{equation}\label{fr}
  f(r)\ =\ 1-\frac{2M}{r}+\frac{4M}{\sqrt{\pi}r}\:\Gamma\left(\frac{3}{2},\frac{r^2}{4\theta}\right).
\end{equation}
Note that when $r/\sqrt{\theta}\to+\infty$, the term $\Gamma(3/2,r^2/4\theta)$ tends to zero, yielding the classic Schwarzschild metric. If we introduce the following rescaling
\begin{equation}
  x = \frac{r}{2M}, \qquad 
  \mu = \frac{M}{\sqrt{\theta}}
\end{equation}
coupled with the identities \cite{Abramowitz1972}
\begin{equation}\label{ids}
  \gamma\left(\frac{3}{2},w^2\right) = \frac{1}{2}\gamma\left(\frac{1}{2},w^2\right)-we^{-w^2},\quad\gamma\left(\frac{1}{2},w^2\right) = \sqrt{\pi}\mbox{erf}(w),
\end{equation}
we can cast (\ref{fr}) in the form
\begin{equation}\label{f}
  f(x)\ =\ 1 - \frac{\mbox{erf}\left(\mu x\right)}{x}+\frac{2\mu}{\sqrt{\pi}}e^{-\mu^2 x^2},
\end{equation}
where $\mbox{erf}(\cdot)$ represents the error function.  The picture emerging from  Figure~\ref{fig0} where $f$ has been plotted against $x$, is the following: an extreme noncommutative geometry inspired Schwarzschild black hole appears when the rescaled mass takes the critical value $\mu_e = 1.904119076\ldots$. In this regime, the Cauchy horizon $x_c$ and the event horizon $x_h$ coincide, and their values correspond to $x_e = x_c = x_h = 0.7936575898\ldots$ Whenever $\mu < \mu_{e}$, the line element (\ref{metric}) represents a naked gravitational droplet regular at $x = 0$ \cite{Nicolini2006PLB}. Finally, two distinct horizons exist for $\mu > \mu_{e}$. Differently as in the case of the Schwarzschild metric, if we probe into shorter distances, \emph{i.e.} $r \ll \sqrt{\theta}$, it is possible to show the absence of a curvature central singularity which is instead replaced by a regular deSitter core \cite{Nicolini2006PLB}. In the present work, we focus our analysis on the QNMs for the nonextreme and extreme cases.

Considering this, it is advantageous to implement the rescaling $z = x/x_h$, as this approach effectively maps the event horizon to $1$ in both the extreme and nonextreme cases. Hence, the $g_{00}$ metric coefficient, as expressed by (\ref{f}), becomes
\begin{equation}\label{fz}
    f(z)=1-\frac{\mbox{erf}\left(\mu x_h z\right)}{x_h z}+\frac{2\mu}{\sqrt{\pi}}e^{-\mu^2 x_h^2 z^2},
\end{equation}
while in the presence of a noncommutative geometry-inspired Schwarzschild black hole, the equation governing a massless Klein--Gordon field (assumed to have a time dependence of the form $e^{-i\omega t}$ and an angular component described by spherical harmonics) is as follows \cite{Batic2019EPJC}
\begin{equation}\label{ODE01}
    f(r)\frac{d}{dr}\left(f(r)\frac{d\psi_{\omega\ell\epsilon}}{dr}\right)+\left[\omega^2-U_\epsilon(r)\right]\psi_{\omega\ell\epsilon}(r)=0,\qquad
    U_\epsilon(r)=f(r)\left[\frac{\epsilon}{r}\frac{df}{dr}+\frac{\ell(\ell+1)}{r^2}\right],\qquad
    \epsilon=1-s^2
\end{equation}
with $\ell = 0, 1, 2, \ldots$ and $\epsilon = -1$ (massless scalar perturbation $s=0$), $\epsilon = 0$ (electromagnetic perturbation $s=1$), and $\epsilon = 3$ (vector-type gravitational perturbation $s=2$). A comprehensive analysis of the effective potential, along with its graphical representations for both configurations, has been extensively covered in \cite{Batic2019EPJC}. By means of the substitution $r=2Mx_h z$, the above equation can be recast in the equivalent form
\begin{equation}\label{ourODE}
    \frac{f(z)}{4}\frac{d}{dz}\left(f(z)\frac{d\psi_{\Omega\ell\epsilon}}{dz}\right)+\left[x_h^2\Omega^2-V_\epsilon(z)\right]\psi_{\Omega\ell\epsilon}(z)=0,\qquad
    V_\epsilon(z)=\frac{f(z)}{4}\left[\frac{\epsilon}{z}\frac{df}{dz}+\frac{\ell(\ell+1)}{z^2}\right],\qquad
    \Omega=M\omega
\end{equation} 
with $f(z)$ as given in (\ref{fz}). As a consistency check, note that in the limit corresponding to the classic Schwarzschild case, \emph{i.e.} $x_h \to 1$ and $r/\sqrt{\theta} \to \infty$, the above equation correctly reproduces equation $(2)$ in \cite{Leaver1985PRSLA}, where the term $\omega^3$ should be corrected to $\omega^2$. In the following analysis, we focus on computing the QNMs for the spectral problem stated in (\ref{ourODE}). For this purpose, we represent $\Omega$ as $\Omega = \Omega_R + i\Omega_I$, where $\Omega_I < 0$ ensures that the perturbation is dumped in time. The boundary conditions are set so that the radial field exhibits inward radiation at the event horizon and outward radiation at spatial infinity. This necessitates a detailed examination of the solution asymptotic behaviour in (\ref{ourODE}), both near the event horizon ($z \to 1^{+}$) and at large spatial distances ($z \to +\infty$). Moreover, to compute the QNMs using the spectral method, we must recast the differential equation in (\ref{ourODE}) and the appropriate boundary conditions over the compact interval $[-1,1]$. This adjustment is essential as the method expands the regular part of the eigenfunctions by means of Chebyshev polynomials. 

\subsection{The non-extreme case}

This scenario focuses on mass parameters $\mu > \mu_e$. The metric coefficient $g_{00}$ is given as defined in (\ref{fz}). It is important to note that the Cauchy and event horizons are distinct in the present case, and therefore, $g_{00}$ exhibits a simple zero at $z=1$. To establish the QNM boundary conditions at the event horizon and at infinity, we first need to determine the asymptotic behaviour of the radial solution $\psi_{\Omega\ell\epsilon}$ as $z \to 1^{+}$ and as $z \to +\infty$. We can then extract the QNM boundary conditions from this asymptotic data. As a consistency check, we will also verify that in the limit $r/\sqrt{\theta}$, these conditions correctly reproduce the corresponding ones for the classic Schwarzschild case specified in equation $(4)$ of \cite{Leaver1985PRSLA}. We split our analysis by examining the behaviour of the radial field in two different regions. 
\begin{enumerate}
    \item 
    {\underline{Asymptotic behaviour as $z\to 1^+$}}: Given that $z=1$ is a simple zero of $f(z)$, we can represent the latter in the form $f(z) = (z-1) g(z)$ where $g$ is an analytic function at $z=1$ with the property that $g(1) = f'(1) \neq 0$. Here, the prime symbol stands for differentiation with respect to $z$. This representation enables us to reformulate (\ref{ourODE}) in the form
    \begin{eqnarray}
        &&\frac{d^2\psi_{\Omega\ell\epsilon}}{dz^2}+p(z)\frac{d\psi_{\Omega\ell\epsilon}}{dz}+q(z)\psi_{\Omega\ell\epsilon}(z)=0,\label{ODEZ}\\
        &&p(z)=\frac{f^{'}(z)}{(z-1)g(z)},\quad
        q(z)=\frac{4x_h^2\Omega^2}{(z-1)^2 g^2(z)}-\frac{1}{(z-1)g(z)}\left[\frac{\epsilon}{z}f^{'}(z)+\frac{\ell(\ell+1)}{z^2}\right].
    \end{eqnarray}
    Since $p$ and $q$ have poles of order one and two at $z = 1$, respectively, this point is classified as a regular singular point of (\ref{ODEZ}), according to Frobenius theory \cite{Ince1956}. Hence, we can construct solutions of the form
    \begin{equation}
        \psi_{\Omega\ell\epsilon}(z) = (z-1)^\rho\sum_{\kappa=0}^\infty a_\kappa(z-1)^\kappa.
    \end{equation}
    The leading behavior at $z = 1$ is represented by the term $(z-1)^\rho$ where $\rho$ is determined by the indicial equation
    \begin{equation}\label{indicial}
        \rho(\rho-1) + P_0\rho + Q_0 = 0
    \end{equation}
    with
    \begin{equation}
        P_0 = \lim_{z \to 1}(z-1)p(z) = 1, \qquad
        Q_0=\lim_{z \to 1}(z-1)^2 q(z) = \left(\frac{2x_h\Omega}{f^{'}(1)}\right)^2.
    \end{equation}
    The roots of (\ref{indicial}) are $\rho_\pm = \pm 2i x_h\Omega/f^{'}(1)$ and the correct QNM boundary condition at $z = 1$ reads
    \begin{equation}\label{QNMBCz1}
    \psi_{\Omega\ell\epsilon}\underset{{z\to 1^+}}{\longrightarrow} (z-1)^{-2i x_h\alpha\Omega},\quad\alpha=\frac{1}{f^{'}(1)}.
    \end{equation}
    We can easily obtain a compact formula for $f^{'}(1)$ if we observe that 
    \begin{equation}\label{fp1}
        f^{'}(1) = \frac{\mbox{erf}(\mu x_h)}{x_h}-\frac{2\mu}{\sqrt{\pi}}(1+2\mu^2 x_h^2)e^{-\mu^2 x_h^2}
    \end{equation}
    can be further simplified by taking into account the condition $f(1) = 0$. Solving the latter with respect to the error function leads to
    \begin{equation}
        \mbox{erf}(\mu x_h) = x_h\left(1+\frac{2\mu}{\sqrt{\pi}}e^{-\mu^2 x_h^2}\right),
    \end{equation}
    which in turns allows to express (\ref{fp1}) as follows
    \begin{equation}
        f^{'}(1) = 1 - \frac{4\mu^3 x_h^2}{\sqrt{\pi}}e^{-\mu^2 x_h^2}.
    \end{equation}
    At this point, a remark is in order. We note that as $\sqrt{\theta} \to 0$, the  rescaled mass parameter $\mu = M/\sqrt{\theta} \to +\infty$. This implies that $f'(1) \to 1$ aligns with the expected behaviour in the classic Schwarzschild case. Applying this observation to (\ref{QNMBCz1}) and considering that in this limit $x_h \to 1$, it becomes evident that the prescription (\ref{QNMBCz1}) accurately reflects the QNM boundary condition at the event horizon, as delineated in equation $(4)$ of \cite{Leaver1985PRSLA}
\item
{\underline{Asymptotic behaviour as $z\to+\infty$}}: We first rewrite (\ref{ourODE}) as
\begin{equation}\label{ODEas}
        \frac{d^2\psi_{\Omega\ell\epsilon}}{dz^2} + P(z)\frac{d\psi_{\Omega\ell\epsilon}}{dz} + Q(z)\psi_{\Omega\ell\epsilon}(z) = 0, \qquad
        P(z) = \frac{f^{'}(z)}{f(z)}, \qquad
        Q(z) = \frac{4\left[x_h^2\Omega^2-V_\epsilon(z)\right]}{f^2(z)}
    \end{equation}
with $V_\epsilon(z)$ defined in (\ref{ourODE}). In order to classify the point at infinity, it is convenient to introduce the transformation $\tau = 1/z$. Then, the above differential equation becomes
\begin{equation}
    \frac{d^2\psi_{\Omega\ell\epsilon}}{d\tau^2}+C(\tau)\frac{d\psi_{\Omega\ell\epsilon}}{d\tau}+D(\tau)\psi_{\Omega\ell\epsilon}(\tau)=0,\quad
        C(\tau)=\frac{2}{\tau}+\frac{1}{f(\tau)}\frac{df}{d\tau},\quad
        D(\tau)=\frac{4\left[x_h^2\Omega^2-V_\epsilon(\tau)\right]}{\tau^4 f^2(\tau)}.
\end{equation}
Since for $k = 1$ we have
\begin{eqnarray}
    \tau^{k+1}C(\tau) &=& C_0 + C_1\tau + \mathcal{O}(\tau^2), \qquad C_0 = 0, \qquad C_1 = 2,\\
    \tau^{2k+2}D(\tau) &=& D_0 + D_1\tau + \mathcal{O}(\tau^2), \qquad D_0 = 4x_h^2\Omega^2, \quad D_1 = 8x_h\Omega^2
\end{eqnarray}
with $D_0\neq 0$, according to \cite{Bender1999}, the point at infinity is an irregular singular point of rank one. Consequently, the asymptotic behaviour of the solutions to equation (\ref{ODEas}) can be efficiently deduced using the method outlined in \cite{Olver1994MAA}. For this purpose, we start by observing that
\begin{equation}
    P(z) = \sum_{\kappa=0}^\infty\frac{\mathfrak{f}_\kappa}{z^k} = \mathcal{O}\left(\frac{1}{z^2}\right), \qquad
    Q(z) = \sum_{\kappa=0}^\infty\frac{\mathfrak{g}_\kappa}{z^k}=4x_h^2\Omega^2+\frac{8x_h\Omega^2}{z}+\mathcal{O}\left(\frac{1}{z^2}\right).
\end{equation}
Given that at least one of the coefficients $\mathfrak{f}_0$, $\mathfrak{g}_0$, $\mathfrak{g}_1$ is nonzero, a formal solution to (\ref{ODEas}) is represented by \cite{Olver1994MAA}
\begin{equation}\label{olvers}
    \psi^{(j)}_{\Omega\ell\epsilon}(z) = z^{\mu_j}e^{\lambda_j z}\sum_{\kappa=0}^\infty\frac{a_{\kappa,j}}{z^\kappa}, \qquad j \in \{1,2\},
\end{equation}
where $\lambda_1$, $\lambda_2$, $\mu_1$ and $\mu_2$ are the roots of the characteristic equations
\begin{equation}\label{chareqns}
   \lambda^2+\mathfrak{f}_0\lambda+\mathfrak{g}_0=0,\quad
   \mu_j=-\frac{\mathfrak{f}_1\lambda_j+\mathfrak{g}_1}{\mathfrak{f}_0+2\lambda_j}.
\end{equation}
A straightforward computation shows that $\lambda_\pm = \pm 2ix_h\Omega$ and $\mu_\pm = \pm 2i\Omega$. As a result, the QNM boundary condition at space-like infinity can be expressed, as
\begin{equation}\label{QNMBCzinf}
    \psi_{\Omega\ell\epsilon}\underset{{z\to +\infty}}{\longrightarrow} z^{2i\Omega}e^{2i x_h\Omega z}.
\end{equation}
It is gratifying to observe that it reproduces correctly the corresponding condition for the classic Schwarzschild case.
\end{enumerate}
At this point, we can proceed as in \cite{Leaver1985PRSLA}, that is, we transform the radial function $\psi_{\Omega\ell\epsilon}(z)$ into a new radial function $\Phi_{\Omega\ell\epsilon}(z)$ such that the QNM boundary conditions are automatically implemented and  $\Phi_{\sigma\ell\epsilon}(z)$ is regular at $z = 1$ and at space-like infinity. To this aim, we consider the transformation
\begin{equation}\label{Ansatz}
    \psi_{\Omega\ell\epsilon}(z) = z^{2i(1+x_h\alpha)\Omega}(z-1)^{-2i x_h\alpha\Omega}e^{2ix_h \Omega(z-1)} \Phi_{\Omega\ell\epsilon}(z),\quad
    \alpha=\frac{1}{1-\frac{4\mu^3 x_h^2}{\sqrt{\pi}}e^{-\mu^2 x_h^2}}.
\end{equation}
Note that in the limit of a vanishing noncommutative parameter, the above ansatz reduces to the corresponding one represented by equation (5) in \cite{Leaver1985PRSLA} for the classic Schwarzschild case. If we substitute (\ref{Ansatz}) into (\ref{ourODE}), we end up with the following ordinary differential equation for the radial eigenfunctions, namely
\begin{equation}\label{ODEznone}
    P_2(z)\Phi^{''}_{\Omega\ell\epsilon}(z) + P_1(z)\Phi^{'}_{\Omega\ell\epsilon}(z) + P_0(z)\Phi_{\Omega\ell\epsilon}(z) = 0
\end{equation}
with
\begin{eqnarray}
    P_2(z)&=&\frac{z^2(z-1)^2}{4}f^2(z),\\
    P_1(z)&=&z(z-1)f(z)\left\{\frac{z(z-1)}{4}f^{'}(z)-i\Omega f(z)\left[1+x_h\alpha+(x_h-1)z-x_h z^2\right]\right\},\\
    P_0(z)&=&-\Omega^2 Q_+(z)Q_{-}(z)+i\Omega f(z)L(z)-z^2(z-1)^2 V_\epsilon(z),\\
    Q_\pm(z)&=&x_h z(z-1)\left[f(z)\pm 1\right]-f(z)\left(1+x_h\alpha-z\right),\\
    L(z)&=&\frac{(z-1)^2}{2}\left[z(1+x_h z)f^{'}(z)-f(z)\right]-\frac{x_h\alpha}{2}\left[z(z-1)f^{'}(z)+(1-2z)f(z)\right].
\end{eqnarray}
Let us now introduce the transformation $z = 2/(1-y)$ mapping the point at infinity and the event horizon to $y = 1$ and $y = -1$, respectively. Furthermore, a dot denotes differentiation with respect to the new variable $y$. Then, equation (\ref{ODEznone}) becomes
\begin{equation}\label{ODEynone}
    S_2(y)\ddot{\Phi}_{\Omega\ell\epsilon}(y) + S_1(y)\dot{\Phi}_{\Omega\ell\epsilon}(y) + S_0(y)\Phi_{\Omega\ell\epsilon}(y) = 0,
\end{equation}
where
\begin{eqnarray}
  S_2(y) &=& \frac{(1+y)^2}{4} f^2(y), \label{S2onone} \\
  S_1(y) &=& i\Omega\frac{1+y}{(1-y)^2}f^2(y)\left[(1+y)(1+2x_h-y)-x_h\alpha(1-y)^2\right]-\frac{(1+y)^2}{2(1-y)}f^2(y)+\frac{(1+y)^2}{4} f(y)\dot{f}(y), \label{S1onone}\\
  S_0(y) &=& \Omega^2\Sigma_2(y)+i\Omega\Sigma_1(y)+\Sigma_0(y) \label{S0onone}
\end{eqnarray}
with
\begin{eqnarray}
    \Sigma_2(y) &=& \frac{4x_h^2(1+y)^2}{(1-y)^4}-\frac{f^2(y)}{(1-y)^4}\left[(1+y)(1+2x_h-y)-x_h\alpha(1-y)^2\right]^2,\\
    \Sigma_1(y) &=& \frac{f(y)}{2}\left\{
    \left(\frac{1+y}{1-y}\right)^2\left[(1+2x_h-y)\dot{f}(y)-f(y)\right]+\frac{x_h\alpha}{1-y}\left[(3+y)f(y)-(1-y^2)\dot{f}(y)\right]
    \right\},\\
    \Sigma_0(y) &=& -\frac{4(1+y)^2}{(1-y)^4}V_\epsilon(y).
\end{eqnarray}
Notice that we must also require that $\Phi_{\Omega\ell\epsilon}(y)$ is regular at $y=\pm 1$. As a result of the transformation introduced above, we have 
\begin{equation}\label{fv}
  f(y) = 1 - \frac{1-y}{2x_h}\mbox{erf}\left(\frac{2\mu x_h}{1-y}\right) + \frac{2\mu}{\sqrt{\pi}}e^{-\frac{4\mu^2 x_h^2}{(1-y)^2}}, \qquad 
  V_\epsilon(y) = \frac{(1-y)^2}{16}f(y)\left[\epsilon (1-y)\dot{f}(y)+\ell(\ell+1)\right].
\end{equation}

\begin{table}
\caption{Classification of the points $y=\pm 1$ for the relevant functions defined by  (\ref{fv}), (\ref{S2onone}), (\ref{S1onone}) and (\ref{S0onone}). The abbreviations $z$ ord $n$ and $p$ ord $m$ stand for zero of order $n$ and pole of order $m$, respectively.}
\begin{center}
\begin{tabular}{ | c | c | c | c | c | c | c | c }
\hline
$y$  & $f(y)$  & $V_\epsilon(y)$ & $S_2(y)$ & $S_1(y)$ & $S_0(y)$\\ \hline
$-1$ & z \mbox{ord} 1 & z \mbox{ord} 1 & z \mbox{ord} 4& z \mbox{ord} 3 & z \mbox{ord} 3 \\ \hline
$+1$ & $+1$  & z \mbox{ord} 2 & $+1$ & p \mbox{ord} 2 & p \mbox{ord} 2\\ \hline
\end{tabular}
\label{tableEinsnone}
\end{center}
\end{table}

Table~\ref{tableEinsnone} shows that the coefficients of the differential equation (\ref{ODEynone}) share a common zero of order $3$ at $y = -1$ while $y = 1$ is a pole of order $2$ for the coefficients $S_1$ and $S_0$. Hence, in order to apply the spectral method, we need to multiply (\ref{ODEynone}) by $(1-y)^2/(1+y)^3$. As a result, we end up with the following differential equation
\begin{equation}\label{ODEhynone}
    M_2(y)\ddot{\Phi}_{\Omega\ell\epsilon}(y) + M_1(y)\dot{\Phi}_{\Omega\ell\epsilon}(y) + M_0(y)\Phi_{\Omega\ell\epsilon}(y) = 0,
\end{equation}
where
\begin{equation}\label{S210honone}
  M_2(y) = \frac{(1-y)^2}{4(1+y)}f^2(y), \qquad
  M_1(y) = i\Omega N_1(y)+N_0(y), \qquad
  M_0(y) = \Omega^2 C_2(y)+i\Omega C_1(y)+C_0(y)
\end{equation}
with
\begin{eqnarray}
    N_1(y) &=& f^2(y)\left[\frac{1+2x_h-y}{1+y}-x_h\alpha\left(\frac{1-y}{1+y}\right)^2\right], \quad
    N_0(y) = \frac{f(y)}{4(1+y)}\frac{d}{dy}\left((1-y)^2 f(y)\right),\label{N0}\\
    C_2(y) &=& \frac{4x_h^2}{(1+y)(1-y)^2}-\frac{f^2(y)}{(1+y)^3(1-y)^2}\left[(1+y)(1+2x_h-y)-x_h\alpha(1-y)^2\right]^2,\label{C2}\\
    C_1(y) &=& \frac{f(y)}{2(1+y)}\left\{(1+2x_h-y)\dot{f}(y)-f(y)+x_h\alpha\frac{1-y}{(1+y)^2}\left[(3+y)f(y)-(1-y^2)\dot{f}(y)\right]\right\},\label{C1}\\
    C_0(y) &=& -\frac{4V_\epsilon(y)}{(1+y)(1-y)^2}.\label{C0}
\end{eqnarray}
It can be easily verified with Maple that
\begin{eqnarray}
    &&\lim_{y\to 1^{-}}M_2(y)=0=\lim_{y\to -1^{+}}M_2(y),\\
    &&\lim_{y\to 1^{-}}M_1(y)=ix_h\Omega,\quad
    \lim_{y\to -1^{+}}M_1(y)=i\Omega\Lambda_1+\Lambda_0,\\
    &&\lim_{y\to 1^{-}}M_0(y)=A_2\Omega^2 +A_0,\quad
     \lim_{y\to -1^{+}}M_0(y)=B_2\Omega^2+i\Omega B_1+B_0,
\end{eqnarray}
where
\begin{eqnarray}
  \Lambda_1 &=& \frac{x_h\left(4\mu^3 x_h^2 e^{-\mu^2 x_h^2}-\sqrt{\pi}\right)}{\sqrt{\pi}},\quad
  \Lambda_0 = \frac{\left(4\mu^3 x_h^2 e^{-\mu^2 x_h^2}-\sqrt{\pi}\right)^2}{4\pi},\\
  A_2 &=& 1-\frac{\sqrt{\pi}x_h^2}{4\mu^3 x_h^2 e^{-\mu^2 x_h^2}-\sqrt{\pi}},\quad A_0=-\frac{\ell(\ell+1)}{8},\label{Acoefnone}\\
  B_2 &=& -\frac{x_h(1+x_h)}{\sqrt{\pi}}(4\mu^3 x_h^2 e^{-\mu^2 x_h^2}-\sqrt{\pi})+\frac{2x_h^2\left[2\mu^3 x_h^2(1+\mu^2 x_h^2)e^{-\mu^2 x_h^2}-\sqrt{\pi}\right]}{4\mu^3 x_h^2 e^{-\mu^2 x_h^2}-\sqrt{\pi}},\label{Dcoefnone}\\
  B_1 &=& \frac{\mathfrak{a}_3 e^{-3\mu^2 x_h^2}+\mathfrak{a}_2 e^{-2\mu^2 x_h^2}+\mathfrak{a}_1 e^{-\mu^2 x_h^2}+\mathfrak{a}_0}{4\pi(4\mu^3 x_h^2 e^{-\mu^2 x_h^2}-\sqrt{\pi})},\\
  B_0 &=& \frac{4\mu^3 x_h^2 e^{-\mu^2 x_h^2}-\sqrt{\pi}}{8\sqrt{\pi}}\left[\ell(\ell+1)-\frac{\epsilon}{\sqrt{\pi}}(4\mu^3 x_h^2 e^{-\mu^2 x_h^2}-\sqrt{\pi})\right]\label{B0}
\end{eqnarray}
and
\begin{eqnarray}
    \mathfrak{a}_3 &=& 64\mu^9 x_h^6(1+x_h),\quad
    \mathfrak{a}_2 = -16\sqrt{\pi}\mu^6 x_h^4(\mu^2 x_h^3+4x_h+3),\\
    \mathfrak{a}_1 &=& 4\pi\mu^3 x_h^2(\mu^2 x_h^3+6x_h+3),\quad \mathfrak{a}_0=-\pi^{3/2}(1+3x_h).
\end{eqnarray}
In the final step leading to the application of the spectral method, we recast the differential equation (\ref{ODEhynone}) into the following form
\begin{equation}\label{TSCH}
  \widehat{L}_0\left[\Phi_{\Omega\ell\epsilon}, \dot{\Phi}_{\Omega\ell\epsilon}, \ddot{\Phi}_{\Omega\ell\epsilon}\right] +  i\widehat{L}_1\left[\Phi_{\Omega\ell\epsilon}, \dot{\Phi}_{\Omega\ell\epsilon}, \ddot{\Phi}_{\Omega\ell\epsilon}\right]\Omega +  \widehat{L}_2\left[\Phi_{\Omega\ell\epsilon}, \dot{\Phi}_{\Omega\ell\epsilon}, \ddot{\Phi}_{\Omega\ell\epsilon}\right]\Omega^2 = 0
\end{equation}
with
\begin{eqnarray}
  \widehat{L}_0\left[\Phi_{\Omega\ell\epsilon}, \dot{\Phi}_{\Omega\ell\epsilon}, \ddot{\Phi}_{\Omega\ell\epsilon}\right] &=& \widehat{L}_{00}(y)\Phi_{\Omega\ell\epsilon} + \widehat{L}_{01}(y)\dot{\Phi}_{\Omega\ell\epsilon} + \widehat{L}_{02}(y)\ddot{\Phi}_{\Omega\ell\epsilon},\label{L0none}\\
  \widehat{L}_1\left[\Phi_{\Omega\ell\epsilon}, \dot{\Phi}_{\Omega\ell\epsilon}, \ddot{\Phi}_{\Omega\ell\epsilon}\right] &=& \widehat{L}_{10}(y)\Phi_{\Omega\ell\epsilon} + \widehat{L}_{11}(y)\dot{\Phi}_{\Omega\ell\epsilon} + \widehat{L}_{12}(y)\ddot{\Phi}_{\Omega\ell\epsilon}, \label{L1none}\\
  \widehat{L}_2\left[\Phi_{\Omega\ell\epsilon}, \dot{\Phi}_{\Omega\ell\epsilon}, \ddot{\Phi}_{\Omega\ell\epsilon}\right] &=& \widehat{L}_{20}(y)\Phi_{\Omega\ell\epsilon} + \widehat{L}_{21}(y)\dot{\Phi}_{\Omega\ell\epsilon} + \widehat{L}_{22}(y)\ddot{\Phi}_{\Omega\ell\epsilon}.\label{L2none}
\end{eqnarray}
Moreover, in Table~\ref{tableZweinone}, we have summarized the $\widehat{L}_{ij}$ appearing in (\ref{L0none})--(\ref{L2none}) and their limiting values at $y = \pm 1$.

\begin{table}
\caption{Definitions of the coefficients $\widehat{L}_{ij}$ and their corresponding behaviours at the endpoints of the interval $-1 \leq y \leq 1$. The symbols appearing in this table have been defined in (\ref{S210honone})-(\ref{B0}).}
\begin{center}
\begin{tabular}{ | c | c | c | c | c | c | c | c }
\hline
$(i,j)$  & $\displaystyle{\lim_{y\to -1^+}}\widehat{L}_{ij}$  & $\widehat{L}_{ij}$ & $\displaystyle{\lim_{y\to 1^-}}\widehat{L}_{ij}$  \\ \hline
$(0,0)$ &  $B_0$          & $C_0$                  & $A_0$\\ \hline
$(0,1)$ &  $\Lambda_0$    & $N_0$                  & $0$\\ \hline
$(0,2)$ &  $0$            & $M_2$                  & $0$\\ \hline 
$(1,0)$ &  $B_1$          & $C_1$                  & $0$\\ \hline 
$(1,1)$ &  $\Lambda_1$    & $N_1$                  & $x_h$\\ \hline 
$(1,2)$ &  $0$            & $0$                    & $0$\\ \hline 
$(2,0)$ &  $B_2$          & $C_2$                  & $A_2$\\ \hline
$(2,1)$ &  $0$            & $0$                    & $0$\\ \hline
$(2,2)$ &  $0$            & $0$                    & $0$\\ \hline
\end{tabular}
\label{tableZweinone}
\end{center}
\end{table} 

\subsection{The extreme case}

In this scenario characterized by $\mu = \mu_e$, we still introduce the rescaling $x=r/(2M)$. However, since the Cauchy and the event horizon coincide, \emph{i.e.} $x_e = x_c = x_h = 0.7936575898\ldots$,  it is useful to introduce a second rescaling given by $\xi = x/x_e$. Hence, the $g_{00}$ metric coefficient, as expressed by (\ref{f}), becomes
\begin{equation}\label{fze}
    f_e(\xi)=1-\frac{\mbox{erf}\left(\mu_e x_e\xi\right)}{x_e\xi}+\frac{2\mu_e}{\sqrt{\pi}}e^{-\mu_e^2 x_e^2\xi^2},
\end{equation}
while the radial part of a massless Klein--Gordon field can be expressed in the equivalent form
\begin{equation}\label{ourODEe}
    \frac{f_e(\xi)}{4}\frac{d}{d\xi}\left(f_e(\xi)\frac{d\psi_{\Omega\ell\epsilon}}{d\xi}\right)+\left[x_e^2\Omega^2-V_\epsilon(\xi)\right]\psi_{\Omega\ell\epsilon}(\xi)=0,\quad
    V_\epsilon(\xi)=\frac{f_e(\xi)}{4}\left[\frac{\epsilon}{\xi}\frac{df_e}{d\xi}+\frac{\ell(\ell+1)}{\xi^2}\right],\quad
    \Omega=M\omega
\end{equation}
with $f_e(\xi)$ as given in (\ref{fze}). The fact that $f_e(\xi)$ exhibits a zero of order two at $\xi = 1$, implies that $f_e(1) = 0 = f^{'}(1)$ where the prime denotes differentiation with respect to $\xi$. This observation allows us to derive the following functional relations
\begin{equation}\label{erf-exp}
    e^{-\mu_e^2 x_e^2}=\frac{\sqrt{\pi}}{4\mu_e^3 x_e^2},\quad
    \mbox{erf}\left(\mu_e x_e\right)=\frac{1+2\mu_e^2 x_e^2}{2\mu_e^2 x_e},
\end{equation}
which, in turn, plays an important role in simplifying the forthcoming computations. In order to derive the QNM boundary conditions at the event horizon and at infinity, we first need to determine the asymptotic behaviour of the radial solution $\psi_{\Omega\ell\epsilon}$ as $\xi \to 1^{+}$ and as $\xi \to +\infty$. We can then accurately extract the QNM boundary conditions from this asymptotic data.
\begin{enumerate}
    \item 
    {\underline{Asymptotic behaviour as $\xi \to 1^+$}}: Taking into account that $\xi = 1$ is a double zero of $f_e(\xi)$, we can represent the latter in the form $f_e(\xi) = (\xi-1)^2 h(\xi)$ where $h$ is an analytic function at $\xi = 1$ with the property that $h(1) = f_e^{''}(1)/2 = \mu_e^2 x_e^2 - 1 \approx 1.2837$. This representation enables us to reformulate (\ref{ourODEe}) in the form
    \begin{eqnarray}
        &&\frac{d^2\psi_{\Omega\ell\epsilon}}{d\xi^2}+\mathfrak{p}(\xi)\frac{d\psi_{\Omega\ell\epsilon}}{d\xi}+\mathfrak{q}(\xi)\psi_{\Omega\ell\epsilon}(\xi)=0,\label{ODEZe}\\
        &&\mathfrak{p}(\xi)=\frac{2}{\xi-1}+\frac{h^{'}(\xi)}{h(\xi)},\quad
        \mathfrak{q}(\xi)=\frac{4x_e^2\Omega^2}{(\xi-1)^4 h^2(\xi)}-\frac{1}{(\xi-1)^2 h(\xi)}\left[\frac{\epsilon}{\xi}f_e^{'}(\xi)+\frac{\ell(\ell+1)}{\xi^2}\right].
    \end{eqnarray}
Because $\mathfrak{q}$ has a fourth-order pole at $\xi = 1$, it follows that $\xi = 1$ qualifies as an irregular singularity, rendering Frobenius's theory inapplicable in this scenario. On the other hand, since for $k = 1$ we have
\begin{equation}
    (\xi-1)^{k+1}\mathfrak{p}(\xi)=\mathcal{O}(\xi-1),\quad 
    (\xi-1)^{2k+2}\mathfrak{q}(\xi)=\mathfrak{d}_0+\mathcal{O}(\xi-1)^2,\quad \mathfrak{d}_0=\frac{4x_e^2\Omega^2}{h^2(1)}
\end{equation}
with $\mathfrak{d}_0 \neq 0$, then, according to \cite{Bender1999}, $\xi = 1$ is an irregular singular point of rank one. Consequently, the leading behaviour of the solutions to equation (\ref{ODEZe}) in a neighbourhood of the event horizon can be efficiently deduced using the method outlined in \cite{Olver1994MAA}. To this purpose, we start by observing that by means of the transformation $\tau = (\xi-1)^{-1}$ mapping the event horizon at infinity and infinity to zero, (\ref{ODEZe}) becomes
\begin{eqnarray}
    &&\frac{d^2\psi_{\Omega\ell\epsilon}}{d\tau^2}+\mathfrak{C}(\tau)\frac{d\psi_{\Omega\ell\epsilon}}{d\tau}+\mathfrak{D}(\tau)\psi_{\Omega\ell\epsilon}(\tau)=0,\\
    &&\mathfrak{C}(\tau)=\sum_{\kappa=0}^\infty\frac{\mathfrak{c}_\kappa}{\tau^k}=\mathcal{O}\left(\frac{1}{\tau^2}\right),\quad
        \mathfrak{D}(\tau)=\sum_{\kappa=0}^\infty\frac{\mathfrak{d}_\kappa}{\tau^k}=\frac{4x_e^2\Omega^2}{(\mu_e^2 x_e^2-1)^2}+\frac{16 x_e^2\Omega^2(\mu_e^4 x_e^4-\mu_e^2 x_e^2-1)}{3(\mu_e^2 x_e^2-1)^3\tau}+\mathcal{O}\left(\frac{1}{\tau^2}\right).
\end{eqnarray}
Since at least one of the coefficients $\mathfrak{c}_0$, $\mathfrak{d}_0$, $\mathfrak{d}_1$ is nonzero, a formal solution to (\ref{ODEZe}) is given by \cite{Olver1994MAA}
\begin{equation}
    \psi^{(\pm)}_{\Omega\ell\epsilon}(\tau)=\tau^{\mu_\pm}e^{\lambda_\pm \tau}\sum_{\kappa=0}^\infty\frac{\mathfrak{a}_{\kappa,\pm}}{\tau^\kappa},
\end{equation}
where $\lambda_\pm$, and $\mu_\pm$ are the roots of the characteristic equations
\begin{equation}
   \lambda_\pm^2+\mathfrak{d}_0=0,\quad
   \mu_\pm=-\frac{\mathfrak{d}_1}{2\lambda_\pm}.
\end{equation}
A straightforward computation shows that
\begin{equation}\label{lmu}
    \lambda_\pm=\pm\frac{2i x_e\Omega}{\mu_e^2 x_e^2-1},\quad
    \mu_\pm=\pm\frac{4i x_e\Omega(\mu_e^4 x_e^4-\mu_e^2 x_e^2-1)}{3(\mu_e^2 x_e^2-1)^2}.
\end{equation}
At this point, it is important to note that a radial field exhibiting solely inward radiation near the event horizon ($\xi \to 1^+$) corresponds, under the transformation $\tau = (\xi-1)^{-1}$, to an outward radiating field as we approach spatial infinity ($\tau \to +\infty^{-}$). This indicates that we should choose the plus sign in the formulas above. Consequently, the correct QNM boundary condition at $\xi = 1$ is as follows
\begin{equation}\label{QNMBCe1}
    \psi_{\Omega\ell\epsilon}\underset{{\xi\to 1^+}}{\longrightarrow} (\xi-1)^{-\mu_+}\mbox{exp}\left(\frac{\lambda_+}{\xi-1}\right)
\end{equation}
with $\mu_+$ and $\lambda_+$ defined in (\ref{lmu}).
\item 
{\underline{Asymptotic behaviour as $\xi \to +\infty$}}: By means of the transformation $\eta = 1/\xi$, it is not difficult to verify that the point at infinity is again an irregular singular point of rank one. Therefore, the asymptotic behaviour of the solutions to equation (\ref{ODEZe}) can be derived according to the method outlined in \cite{Olver1994MAA}. To this purpose, we  observe that
\begin{equation}
    \mathfrak{p}(\xi)=\sum_{\kappa=0}^\infty\frac{\widehat{\mathfrak{f}}_\kappa}{\xi^k}=\mathcal{O}\left(\frac{1}{\xi^2}\right),\quad
    \mathfrak{q}(\xi)=\sum_{\kappa=0}^\infty\frac{\widehat{\mathfrak{g}}_\kappa}{\xi^k}=4x_e^2\Omega^2+\frac{8x_e\Omega^2}{\xi}+\mathcal{O}\left(\frac{1}{\xi^2}\right).
\end{equation}
With the help of (\ref{chareqns}), we immediately find that $\lambda_\pm = \pm 2i x_e\Omega$ and $\mu_\pm = \pm 2i\Omega$. Hence, the QNM boundary condition at space-like infinity can be expressed as
 \begin{equation}\label{QNMBCzinfe}
    \psi_{\Omega\ell\epsilon}\underset{{\xi\to +\infty}}{\longrightarrow} \xi^{2i\Omega}e^{2i x_e\Omega\xi}.
    \end{equation}
\end{enumerate}
Let us transform the radial function $\psi_{\Omega\ell\epsilon}(\xi)$ into a new radial function $\Phi_{\Omega\ell\epsilon}(\xi)$ such that the QNM boundary conditions are automatically implemented and  $\Phi_{\sigma\ell s}(\xi)$ is regular at $\xi = 1$ and at space-like infinity. To this aim, we consider the transformation
\begin{equation}\label{Ansatze}
    \psi_{\Omega\ell\epsilon}(\xi)=\xi^{2i\Omega+\mu_+}(\xi-1)^{-\mu_+}e^{2ix_e \Omega(\xi-1)+\frac{\lambda_+}{\xi-1}} \Phi_{\Omega\ell\epsilon}(\xi).
\end{equation}
If we rewrite it in a more compact form, namely
\begin{eqnarray}
    \psi_{\Omega\ell\epsilon}(\xi)&=&\xi^{2ia\Omega}(\xi-1)^{-2i(a-1)\Omega}e^{2ix_e\Omega\eta(\xi)} \Phi_{\Omega\ell\epsilon}(\xi),\\
    \eta(\xi)&=&\xi-1+\frac{1}{(\mu_e^2 x_e^2-1)(\xi-1)},\quad
    a=1+\frac{2x_e(\mu_e^4 x_e^4-\mu_e^2 x_e^2-1)}{3(\mu_e^2 x_e^2-1)^2},
\end{eqnarray}
and we replace it into (\ref{ourODEe}), we end up with the differential equation
\begin{equation}\label{ODEzext}
    P_{2e}(\xi)\Phi^{''}_{\Omega\ell\epsilon}(\xi)+P_{1e}(\xi)\Phi^{'}_{\Omega\ell\epsilon}(\xi)+P_{0e}(\xi)\Phi_{\Omega\ell\epsilon}(\xi)=0
\end{equation}
with
\begin{eqnarray}
    P_{2e}(\xi)&=&\frac{\xi^2(\xi-1)^2}{4}f^2_e(\xi),\\
    P_{1e}(\xi)&=&\xi(\xi-1)f_e(\xi)\left\{\frac{\xi(\xi-1)}{4}f^{'}_e(\xi)+i\Omega f_e(\xi)\left[x_e\xi(\xi-1)\eta^{'}(\xi)+\xi-a\right]\right\},\\
    P_{0e}(\xi)&=&-\mathfrak{Q}_+(\xi)\mathfrak{Q}_-(\xi)\Omega^2+i\Omega f_e(\xi)\mathfrak{L}(\xi)-\xi^2(\xi-1)^2 V_\epsilon(\xi),\\
    \mathfrak{Q}_\pm(\xi)&=&x_e \xi(\xi-1)f_e(\xi)\eta^{'}(\xi)\pm x_e\xi(\xi-1)+(\xi-a)f_e(\xi),\\
    \mathfrak{L}(\xi)&=&\frac{x_e}{2}\xi^2(\xi-1)^2\left[f_e(\xi)\eta^{'}(\xi)\right]^{'}+\frac{1}{2}\xi(\xi-1)(\xi-a)f_e^{'}(\xi)-\frac{1}{2}(\xi^2-2a\xi+a)f_e(\xi).
\end{eqnarray}
Let us introduce the transformation $\xi = 2/(1-y)$ mapping the point at infinity and the event horizon to $y = 1$ and $y = -1$, respectively. Furthermore, a dot denotes differentiation with respect to the new variable $y$. Then, equation (\ref{ODEzext}) becomes
\begin{equation}\label{ODEye}
    S_{2e}(y)\ddot{\Phi}_{\Omega\ell\epsilon}(y)+S_{1e}(y)\dot{\Phi}_{\Omega\ell\epsilon}(y)+S_{0e}(y)\Phi_{\Omega\ell\epsilon}(y)=0,
\end{equation}
where
\begin{eqnarray}
    S_{2e}(y)&=&\frac{(1+y)^2}{4} f_e^2(y),\label{S2oe}\\
    S_{1e}(y)&=&i\Omega\frac{1+y}{1-y}f_e^2(y)\left[x_e(1-y^2)\dot{\eta}(y)+2-a(1-y)\right]-\frac{(1+y)^2}{2(1-y)}f_e^2(y)+\frac{(1+y)^2}{4}f_e(y)\dot{f}_e(y),\label{S1oe}\\
    S_{0e}(y)&=&\Omega^2\Sigma_{2e}(y)+i\Omega\Sigma_{1e}(y)+\Sigma_{0e}(y)\label{S0oe}
\end{eqnarray}
with
\begin{eqnarray}
\Sigma_{2e}(y)&=&\frac{4x_e^2(1+y)^2}{(1-y)^4}-\frac{f_e^2(y)}{(1-y)^2}\left[x_e(1-y^2)\dot{\eta}(y)+2-a(1-y)\right]^2,\\
\Sigma_{1e}(y)&=&\frac{x_e}{2}(1+y)^2 f_e(y)\dot{f}_e(y)\dot{\eta}(y)+\frac{x_e}{2}\frac{(1+y)^2}{1-y}f_e^2(y)\left[(1-y)\ddot{\eta}(y)-2\dot{\eta}(y)\right]+\nonumber\\
&&\frac{1+y}{2(1-y)}\left[2-a(1-y)\right]f_e(y)\dot{f}_e(y)-\frac{f_e^2(y)}{(1-y)^2}\left[2-2a(1-y)+\frac{a}{2}(1-y)^2\right],\\
\Sigma_{1e}(y)&=&-\frac{4(1+y)^2}{(1-y)^4}V_\epsilon(y).
\end{eqnarray}
and the requirement that $\Phi_{\Omega\ell\epsilon}(y)$ is regular at $y = \pm 1$. As a result of the transformation introduced above, we have 
\begin{equation}\label{fve}
  f_e(y) = 1 - \frac{1-y}{2x_e}\mbox{erf}\left(\frac{2\mu_e x_e}{1-y}\right)+\frac{2\mu_e}{\sqrt{\pi}}e^{-\frac{4\mu_e^2 x_e^2}{(1-y)^2}}, \qquad 
  V_\epsilon(y) = \frac{(1-y)^2}{16}f_e(y)\left[\epsilon (1-y)\dot{f}_e(y)+\ell(\ell+1)\right]
\end{equation}
and
\begin{equation}
    \eta(y) = \frac{1+y}{1-y}+\frac{1-y}{(\mu_e^2 x_e^2-1)(1+y)}.
\end{equation}

\begin{table}
\caption{Classification of the points $y = \pm 1$ for the relevant functions entering in (\ref{S2oe}), (\ref{S1oe}) and (\ref{S0oe}). The abbreviations $z$ ord $n$ and $p$ ord $m$ stand for zero of order $n$ and pole of order $m$, respectively.}
\begin{center}
\begin{tabular}{ | l | l | l | l |l |l | l | l}
\hline
$y$  & $f_e(y)$  & $V_\epsilon(y)$ & $\eta(y)$ & $S_{2e}(y)$ & $S_{1e}(y)$ & $S_{0e}(y)$\\ \hline
$-1$ & z \mbox{ord} 2 & z \mbox{ord} 2 & p \mbox{ord} 1 & z \mbox{ord} 6& z \mbox{ord} 4 & z \mbox{ord} 4 \\ \hline
$+1$ & $+1$  & z \mbox{ord} 2 & p \mbox{ord} 1 & $+1$ & p \mbox{ord} 2 & p \mbox{ord} 2\\ \hline
\end{tabular}
\label{table3}
\end{center}
\end{table}
Table~\ref{table3} shows that the coefficients of the  differential equation (\ref{ODEye}) share a common zero of order $4$ at $y = -1$ while $y = 1$ is a pole of order $2$ for the coefficients $S_{1e}(y)$ and $S_{0e}(y)$. Hence, in order to apply the spectral method, we need to multiply (\ref{ODEye}) by $(1-y)^2/(1+y)^4$. As a result, we end up with the following differential equation
\begin{equation}\label{ODEhynonee}
    M_{2e}(y)\ddot{\Phi}_{\Omega\ell\epsilon}(y)+M_{1e}(y)\dot{\Phi}_{\Omega\ell\epsilon}(y)+M_{0e}(y)\Phi_{\Omega\ell\epsilon}(y)=0,
\end{equation}
where
\begin{equation}\label{S210hononee}
  M_{2e}(y)=\frac{(1-y)^2}{4(1+y)^2}f_e^2(y),\quad
  M_{1e}(y)=i\Omega N_{1e}(y)+N_{0e}(y),\quad
  M_{0e}(y)=\Omega^2 C_{2e}(y)+i\Omega C_{1e}(y)+C_{0e}(y)
\end{equation}
with
\begin{eqnarray}
    N_{1e}(y)&=&\frac{1-y}{(1+y)^3}f_e^2(y)\left[x_e(1-y^2)\dot{\eta}(y)+2-a(1-y)\right],\quad
    N_{0e}(y)=\frac{f_e(y)}{4(1+y)^2}\frac{d}{dy}\left((1-y)^2 f_e(y)\right),\label{N0e}\\
    C_{2e}(y)&=&\frac{4x_e^2}{(1-y^2)^2}-\frac{f_e^2(y)}{(1+y)^4}\left[x_e(1-y^2)\dot{\eta}(y)+2-a(1-y)\right]^2,\label{C2e}\\
    C_{1e}(y)&=&\frac{x_e}{2}\left(\frac{1-y}{1+y}\right)^2 f_e(y)\dot{f}_e(y)\dot{\eta}(y)+\frac{x_e}{2}\frac{1-y}{(1+y)^2}f_e^2(y)\left[(1-y)\ddot{\eta}(y)-2\dot{\eta}(y)\right]+\nonumber\\
    &&\frac{1-y}{2(1+y)^3}\left[2-a(1-y)\right]f_e(y)\dot{f}_e(y)-\frac{f_e^2(y)}{(1+y)^4}\left[2-2a(1-y)+\frac{a}{2}(1-y)^2\right],\label{C1e}\\
    C_{0e}(y)&=&-\frac{4V_\epsilon(y)}{(1-y^2)^2}.\label{C0e}
\end{eqnarray}
It can be easily checked with Maple that
\begin{eqnarray}
    &&\lim_{y\to 1^{-}}M_{2e}(y)=0=\lim_{y\to -1^{+}}M_{2e}(y),\quad \lim_{y\to 1^{-}}M_{1e}(y)=\frac{1}{2}ix_e\Omega,\quad
    \lim_{y\to -1^{+}}M_{1e}(y)=i\Omega\Lambda_{1e},\\
    &&\lim_{y\to 1^{-}}M_{0e}(y)=A_{2e}\Omega^2 +A_{0e},\quad
     \lim_{y\to -1^{+}}M_{0e}(y)=B_{2e}\Omega^2+B_{0e},
\end{eqnarray}
where
\begin{eqnarray}
\Lambda_{1e}&=&-\frac{x_e}{2}\left(\mu_e^2 x_e^2-1\right),\quad
A_{2e}=\frac{2\mu_e^4 x_e^6+3\mu_e^4 x_e^4+\mu_e^2 x_e^4-6\mu_e^2 x_e^2-5x_e^2+3}{6\left(\mu_e^2 x_e^2-1\right)^2},\quad A_{0e}=-\frac{\ell(\ell+1)}{16},\label{Acoefnonee}\\
B_{2e}&=&\frac{x_e\left(2\mu_e^8 x_e^9+35\mu_e^6 x_e^7+18\mu_e^6 x_e^6-107\mu_e^4 x_e^5-54\mu_e^4 x_e^4+88\mu_e^2 x_e^3+54\mu_e^2 x_e^2-10x_e-18\right)}{36\left(\mu_e^2 x_e^2-1\right)^2},\label{Dcoefnonee}\\
B_{0e}&=&-\frac{\mu_e^2 x_e^2-1}{16}\ell(\ell+1).\label{B0e}
\end{eqnarray}
Finally, in order to apply the spectral method, we rewrite the differential equation (\ref{ODEhynonee}) into the following form
\begin{equation}\label{TSCHe}
\widehat{L}^{(e)}_0\left[\Phi_{\Omega\ell\epsilon},\dot{\Phi}_{\Omega\ell\epsilon},\ddot{\Phi}_{\Omega\ell\epsilon}\right]+ i\widehat{L}^{(e)}_1\left[\Phi_{\Omega\ell\epsilon},\dot{\Phi}_{\Omega\ell\epsilon},\ddot{\Phi}_{\Omega\ell\epsilon}\right]\Omega+ \widehat{L}_2^{(e)}\left[\Phi_{\Omega\ell\epsilon},\dot{\Phi}_{\Omega\ell\epsilon},\ddot{\Phi}_{\Omega\ell\epsilon}\right]\Omega^2=0
\end{equation}
with
\begin{eqnarray}
\widehat{L}^{(e)}_0\left[\Phi_{\Omega\ell\epsilon},\dot{\Phi}_{\Omega\ell\epsilon},\ddot{\Phi}_{\Omega\ell\epsilon}\right]&=&\widehat{L}^{(e)}_{00}(y)\Phi_{\Omega\ell\epsilon}+\widehat{L}^{(e)}_{01}(y)\dot{\Phi}_{\Omega\ell\epsilon}+\widehat{L}^{(e)}_{02}(y)\ddot{\Phi}_{\Omega\ell\epsilon},\label{L0nonee}\\
\widehat{L}^{(e)}_1\left[\Phi_{\Omega\ell\epsilon},\dot{\Phi}_{\Omega\ell\epsilon},\ddot{\Phi}_{\Omega\ell\epsilon}\right]&=&\widehat{L}^{(e)}_{10}(y)\Phi_{\Omega\ell\epsilon}+\widehat{L}^{(e)}_{11}(y)\dot{\Phi}_{\Omega\ell\epsilon}+\widehat{L}^{(e)}_{12}(y)\ddot{\Phi}_{\Omega\ell\epsilon},\label{L1nonee}\\
\widehat{L}^{(e)}_2\left[\Phi_{\Omega\ell\epsilon},\dot{\Phi}_{\Omega\ell\epsilon},\ddot{\Phi}_{\Omega\ell\epsilon}\right]&=&\widehat{L}^{(e)}_{20}(y)\Phi_{\Omega\ell\epsilon}+\widehat{L}^{(e)}_{21}(y)\dot{\Phi}_{\Omega\ell\epsilon}+\widehat{L}^{(e)}_{22}(y)\ddot{\Phi}_{\Omega\ell\epsilon}.\label{L2nonee}
\end{eqnarray}
Moreover, in Table~\ref{table4}, we have summarized the $\widehat{L}^{(e)}_{ij}$ appearing in (\ref{L0nonee})--(\ref{L2nonee}) and their limiting values at $y = \pm 1$.

\begin{table}
\caption{Definitions of the coefficients $\widehat{L}^{(e)}_{ij}$ and their corresponding behaviours at the endpoints of the interval $-1\leq y\leq 1$. The symbols appearing in this table have been defined in (\ref{S210hononee})-(\ref{B0e}).}
\begin{center}
\begin{tabular}{ | l | l | l | l |l |l | l | l}
\hline
$(i,j)$  & $\displaystyle{\lim_{y\to -1^+}}\widehat{L}^{(e)}_{ij}$  & $\widehat{L}^{(e)}_{ij}$ & $\displaystyle{\lim_{y\to 1^-}}\widehat{L}^{(e)}_{ij}$  \\ \hline
$(0,0)$ &  $B_{0e}$       & $C_{0e}$                  & $A_{0e}$\\ \hline
$(0,1)$ &  $0$            & $N_{0e}$                  & $0$\\ \hline
$(0,2)$ &  $0$            & $M_{2e}$                  & $0$\\ \hline 
$(1,0)$ &  $0$            & $C_{1e}$                  & $0$\\ \hline 
$(1,1)$ &  $\Lambda_{1e}$ & $N_{1e}$                  & $x_e/2$\\ \hline 
$(1,2)$ &  $0$            & $0$                       & $0$\\ \hline 
$(2,0)$ &  $B_{2e}$       & $C_{2e}$                  & $A_{2e}$\\ \hline
$(2,1)$ &  $0$            & $0$                       & $0$\\ \hline
$(2,2)$ &  $0$            & $0$                       & $0$\\ \hline
\end{tabular}
\label{table4}
\end{center}
\end{table}

\section{Numerical method}

In order to solve the differential eigenvalue problem \eqref{TSCHe} to determine the QNMs along with the corresponding frequencies $\Omega$, we have to discretize the differential operators $\widehat{L}^{(e)}_{j}[\cdot]$ with $j=1,2,3$ defined in (\ref{L0nonee})-(\ref{L2nonee}). Since our problem is posed on the finite interval $[-1, 1]$ without any boundary conditions, more precisely, we only require that the QNMs be regular functions at $y = \pm 1$, then, it is natural to choose a Tchebyshev-type spectral method \cite{Trefethen2000, Boyd2000}. Namely, we are going to expand the function $y \mapsto \Phi_{\Omega\ell\epsilon}(y)$ in the form of a truncated Tchebyshev series
\begin{equation}\label{eq:exp}
\Phi_{\Omega\ell\epsilon}(y)=\sum_{k=0}^{N} a_k T_k(y),
\end{equation}
where $N\ \in\ \mathbb{N}$ is kept as a numerical parameter, $\{a_k\}_{k=0}^{N}\ \subseteq\ \mathds{R}$, and $\{T_k(y)\}_{k=0}^{N}$ are the Tchebyshev polynomials of the first kind
\begin{equation}
    T_k: [-1, 1]\ \longrightarrow\ [-1, 1]\,, \qquad y\ \longmapsto\ \cos\,\bigl(k\arccos y\bigr)\,.
\end{equation}
After substituting expansion \eqref{eq:exp} into the differential equation \eqref{TSCHe}, we obtain an eigenvalue problem with polynomial coefficients. In order to translate it into the realm of numerical linear algebra, we employ the collocation method \cite{Boyd2000}. Specifically, rather than insisting that the polynomial function in \( y \) is identically zero (a condition equivalent to having polynomial solutions for the differential problem as per equation \eqref{TSCHe}), we impose a weaker requirement. This involves ensuring that the polynomial vanishes at \( N+1 \) strategically selected points. The number $N+1$ coincides exactly with the number of unknown coefficients $\{a_k\}_{k=0}^{N}$. For the collocation points, we implemented the Tchebyshev roots grid \cite{Fox1968}
\begin{equation}
  y_k= -\cos{\left(\frac{(2k+1)\pi}{2(n+1)}\right)},\quad k\in\{0, 1,\ldots,N\}.
\end{equation}
In our numerical codes, we also implemented the second option of the Tchebyshev extrema grid
\begin{equation}
  y_k=-\cos{\left(\frac{k\pi}{n}\right)},\quad k\in\{0, 1,\ldots,N\}.
\end{equation}
The users are free to choose their favourite collocation points. Notice that we used the roots grid in our computation, and in any case, the theoretical performance of the two available options is known to be absolutely comparable \cite{Fox1968, Boyd2000}.

Upon implementing the collocation method, we derive a classical matrix-based quadratic eigenvalue problem, as detailed in \cite{Tisseur2001}
\begin{equation}\label{eq:eig}
  (M_0 + iM_1\Omega + M_2\Omega^2)\bf{a} =\bf{0}.
\end{equation}
In this formulation, the square real matrices $M_{j}$, each of size $(N+1)\times(N+1)$ for $j=0,1,2$, represent the spectral discretizations of the operators $\widehat{L}^{(e)}_{j}[\cdot]$, respectively. The problem \eqref{eq:eig} is solved numerically with the \textsc{polyeig} function from \textsc{Matlab}. This polynomial eigenvalue problem yields \(2(N+1)\) potential values for the parameter \(\Omega\). To discern the physical values of \(\Omega\) that correspond to the black hole's QNMs, we first overlap the root plots for various values of \(N\) in equation \eqref{eq:exp}, such as \(N=100, 150, 200\). We then identify the consistent roots whose positions remain stable across these different \(N\) values.

In order to reduce the rounding and other floating point errors, we performed all our computations with multiple precision arithmetic that is built in \textsc{Maple} and which is brought into \textsc{Matlab} by the \textsc{Advanpix} toolbox \cite{mct2015}. All numerical computations reported in this study have been performed with $200$ decimal digits accuracy.

\section{Numerical results}

In this section, we delve into the analysis of the QNMs for both non-extreme and extreme noncommutative geometry-inspired Schwarzschild black holes. The central revelation from our numerical computations is the stability of both black hole regimes under scalar, electromagnetic, and gravitational perturbations. 

The fact that the noncommutative geometry-inspired Schwarzschild black hole reproduces its classical counterpart for large values of the rescaled mass parameter $\mu$ provides a basis for validating our numerical method. By selecting a sufficiently large $\mu$-value, computing the QNMs, and demonstrating their agreement with the results obtained by \cite{Leaver1985PRSLA, IYER1987PRD, Mamani2022EPJC} for the classic Schwarzschild black hole in the $s = 0$ case, we can substantiate the accuracy of our approach. To this end, we refer to Table~\ref{table:0}, where the numerical values listed in the last column were derived using our high-precision computational approach, employing 200 polynomials with an accuracy of $200$ digits.

As an additional note, we would like to highlight that in the extreme case, the transition to the Schwarzschild limit is more complex due to potential non-analyticities in the behavior of the metric and associated physical quantities as \( \theta \) tends towards zero. The extreme configuration in noncommutative geometry, characterized by a specific critical mass \( \mu_e \) at which the inner and outer horizons converge, does not straightforwardly reduce to the Schwarzschild solution by merely setting \( \theta \) to zero. This non-analyticity arises because the extreme condition itself alters the underlying spacetime structure in a way that does not directly map onto the classical Schwarzschild case without noncommutativity.

\begin{table}
\centering
\caption{This table presents the quasinormal frequencies for scalar perturbations ($s = 0$) of the classic Schwarzschild black hole alongside those for a noncommutative geometry-inspired Schwarzschild black hole with a large mass parameter (refer to the last column). The third column showcases numerical values determined by \cite{Leaver1985PRSLA} using the continued fraction method. The fourth column includes third-order WKB results reported by \cite{IYER1987PRD}, and the fifth column presents numerical values from \cite{Mamani2022EPJC}, derived via the spectral method employing 40 polynomials. Here, $\Omega$ represents the dimensionless frequency as defined in equation (\ref{ourODE}). The notation 'N/A' indicates data not available.}
\label{table:0}
 \vspace*{1em}
 \begin{tabular}{||c| c| c| c| c| c||} 
 \hline
 $\ell$ & $n$ & $\Omega$ \cite{Leaver1985PRSLA} & $\Omega$ \cite{IYER1987PRD} & $\Omega$ \cite{Mamani2022EPJC} & $\Omega$, $\mu=10^3$\\ [0.5ex] 
 \hline\hline
 $0$ & $0$ & $0.1105-0.1049i$ & $0.1046-0.1152i$ & $0.1105-0.1049i$ & $0.1104549-0.1048957i$\\ 
     & $1$ & $0.0861-0.3481i$ & $0.0892-0.3550i$ & \mbox{N/A}       & $0.0861169-0.3480525i$\\
 $1$ & $0$ & $0.2929-0.0977i$ & $0.2911-0.0980i$ & $0.2929-0.0977i$ & $0.2929361-0.0976599i$\\
     & $1$ & $0.2645-0.3063i$ & $0.2622-0.3704i$ & $0.2645-0.3063i$ & $0.2644487-0.3062574i$\\
     & $2$ & $0.2295-0.5401i$ & $0.2235-0.5268i$ & \mbox{N/A}       & $0.2295393-0.5401334i$\\ 
     & $3$ & $0.2033-0.7883i$ & $0.1737-0.7486i$ & \mbox{N/A}       & $0.2032584-0.7882978i$\\
 $2$ & $0$ & $0.4836-0.0968i$ & $0.4832-0.0968i$ & $0.4836-0.0968i$ & $0.4836439-0.0967588i$\\
     & $1$ & $0.4639-0.2956i$ & $0.4632-0.2958i$ & $0.4639-0.2956i$ & $0.4638506-0.2956039i$\\
     & $2$ & $0.4305-0.5086i$ & $0.4317-0.5034i$ & $0.4305-0.5086i$ & $0.4305441-0.5085584i$\\
     & $3$ & $0.3939-0.7381i$ & $0.3926-0.7159i$ & \mbox{N/A}       & $0.3938631-0.7380966i$\\ [0.5ex] 
 \hline
 \end{tabular}
\end{table}

It is interesting to observe that \cite{Batic2019EPJC} utilized a sixth-order WKB approximation to show that the WKB method does not converge in critical cases marked by third-order instabilities. These occur for specific choices of the rescaled mass parameters close to $\mu = \mu_e$ for a nonextreme, noncommutative geometry-inspired Schwarzschild black hole. Instabilities were also detected by \cite{Batic2019EPJC} in the case of an extreme black hole. This indicates that conventional methods have exhibited inconsistencies, particularly when applied to non-commutative geometry settings. There are several reasons for this. For instance, the WKB approximation relies on the local properties of the potential near its peak. However, it fails in cases where the second derivative of the potential at the peak is zero. As for the Inverse Potential and Asymptotic Iteration Methods, the former depends on a rough approximation of the potential at the maximum, while the latter depends on the choice of initial parameters or specific local conditions that may not be optimally representative or easily determinable in complex geometries like those modified by the non-commutative parameter. These dependencies can lead to instabilities or divergences in computations, as indicated by the lack of convergence observed in higher-order WKB calculations, suggesting that these methods may reach their analytical limitations in such scenarios.

Unlike local methods, the spectral method utilizes the global properties of the potential, thus avoiding inaccuracies due to local approximations. Moreover, it provides high accuracy and exponential convergence rates for the solutions of differential equations once the irregular part of the solution has been analytically captured. This is particularly beneficial in the quest for QNMs of astrophysical objects. Furthermore, the method is less sensitive to the choice of initial parameters (in contrast with the Asymptotic Iterative Method), making it more robust and reliable for theoretical explorations in gravitational physics where exact solutions are often challenging to compute. For non-commutative geometry-inspired black holes, the spectral method allows us to capture the modified dynamics without the constraints and approximations imposed by traditional techniques. This capability is crucial in revealing that previously reported instabilities are likely artifacts of the limitations of those methods rather than inherent physical instabilities in the black hole models themselves.

Consequently, by harnessing the global analytical capabilities of the spectral method, we have circumvented the shortcomings of traditional approaches like the WKB method. This has allowed us to demonstrate that the instabilities previously reported in non-commutative geometry-inspired black holes are not inherent to the black holes themselves but are rather artifacts resulting from the limitations of local approximation methods.

Therefore, our application of the spectral method provides a clearer, more accurate depiction of the stability characteristics of these black holes. For a comprehensive presentation of our numerical results on massless scalar, electromagnetic, and gravitational perturbations, we invite readers to consult Tables \ref{table:1}, \ref{table:1a}, and \ref{table:1b}. These tables detail the calculated QNMs of a nonextreme noncommutative black hole, illustrating variations across different values of the mass parameter. In contrast, Tables \ref{table:2}, \ref{table:3}, and \ref{table:3b} focus on the extreme case, providing analogous data for the same spectrum of perturbations.

\begin{table}
\centering
\caption{QNMs for scalar perturbations ($s = 0$) of the noncommutative geometry inspired Schwarzschild metric (nonextreme case) for different values of the mass parameter $\mu$. The third and fifth columns represent the numerical values found by \cite{Batic2019EPJC} by means of a sixth-order WKB approximation. The fourth and sixth columns report the corresponding results obtained by means of the spectral method employing $150$ polynomials with an accuracy of $200$ digits. Here, $\Omega$ represents the dimensionless frequency as defined in equation (\ref{ourODE}). The 'SM' abbreviation denotes the 'Spectral Method'.}
\label{table:1}
\vspace*{1em}
\begin{tabular}{||c| c| c| c| c| c||} 
\hline
$\ell$ & 
$n$    & 
$\Omega$, $\mu=1.95$ \cite{Batic2019EPJC} & 
$\Omega$, $\mu=1.95$ (SM) &
$\Omega$, $\mu=2.25$ \cite{Batic2019EPJC} &
$\Omega$, $\mu=2.25$ (SM)\\ [0.5ex] 
\hline\hline
$0$ & $0$ & $0.0397+0.1330i$ & $0.1041-0.0930i$ & $0.0222-0.0944i$ & $0.1101-0.1010i$\\ 
    & $1$ & $0.1555+0.4548i$ & $0.0343-0.3374i$ & $0.0880+0.3947i$ & $0.0595-0.3298i$\\
$1$ & $0$ & $0.2727-0.0818i$ & $0.2890-0.0908i$ & $0.2882-0.3509i$ & $0.2916-0.0958i$\\
    & $1$ & $0.1256-0.2689i$ & $0.2381-0.2807i$ & $0.2449-0.3059i$ & $0.2556-0.2971i$\\
    & $2$ & $0.0779+0.5589i$ & $0.1480-0.5047i$ & $0.1771-0.5284i$ & $0.1961-0.5186i$\\ 
    & $3$ & $0.3072+0.8805i$ & $0.0611-0.7768i$ & $0.0862-0.7594i$ & $0.1056-0.7762i$\\
$2$ & $0$ & $0.4767-0.0889i$ & $0.4806-0.0916i$ & $0.4819-0.0964i$ & $0.4825-0.0957i$\\
    & $1$ & $0.4209-0.2669i$ & $0.4477-0.2667i$ & $0.4561-0.2944i$ & $0.4581-0.2909i$\\
    & $2$ & $0.2999-0.4590i$ & $0.3833-0.4607i$ & $0.4122-0.5015i$ & $0.4134-0.4965i$\\
    & $3$ & $0.1285-0.6908i$ & $0.2925-0.6865i$ & $0.3544-0.7154i$ & $0.3537-0.7160i$
     \\[1ex] 
 \hline
 \end{tabular}
\end{table}

Our findings suggest a remarkable stability of the nonextreme noncommutative Schwarzschild black hole across various perturbations. This observation is particularly noteworthy given the existing literature, such as the study by \cite{Brown2011PLB}, which posited potential instability in similar scenarios. Our results, however, indicate that even in extreme scenarios where the Cauchy horizon coincides with the event horizon, the black hole exhibits stability under the class of perturbations considered. This result extends the work of \cite{Batic2010PLB}, where the stability of the nonextreme noncommutative Schwarzschild black hole's interior was established by investigating the propagation of a massless scalar field between the two horizons.

\begin{table}
\centering
\caption{QNMs for electromagnetic perturbations ($s = 1$) of the noncommutative geometry inspired Schwarzschild metric (nonextreme case) for different values of the mass parameter $\mu$. The corresponding numerical results are obtained by means of the spectral method employing $150$ polynomials with an accuracy of $200$ digits. Here, $\Omega$ represents the dimensionless frequency as defined in equation (\ref{ourODE}).} 
\label{table:1a}
\vspace*{1em}
\begin{tabular}{||c| c| c| c||} 
\hline
$\ell$ & 
$n$ & 
$\Omega$, $\mu=1.95$& 
$\Omega$, $\mu=2.25$ \\ [0.5ex] 
\hline\hline
$1$ & $0$ & $0.2469-0.0831i$ & $0.2478-0.0899i$ \\
    & $1$ & $0.1945-0.2569i$ & $0.2090-0.2810i$ \\
    & $2$ & $0.0965-0.4661i$ & $0.1491-0.4936i$ \\
    & $3$ & $0.0181-0.6972i$ & $0.0656-0.7133i$ \\
$2$ & $0$ & $0.4552-0.0887i$ & $0.4566-0.0936i$ \\
    & $1$ & $0.4220-0.2682i$ & $0.4314-0.2848i$ \\
    & $2$ & $0.3566-0.4561i$ & $0.3855-0.4866i$\\
    & $3$ & $0.2637-0.6656i$ & $0.3254-0.7022i$ 
     \\[1ex] 
 \hline
 \end{tabular}
\end{table}

\begin{table}
\centering
\caption{QNMs for vector-type gravitational perturbations ($s = 2$) of the noncommutative geometry inspired Schwarzschild metric (nonextreme case) for different values of the mass parameter $\mu$. The corresponding numerical results are obtained by means of the spectral method employing $150$ polynomials with an accuracy of $200$ digits. Here, $\Omega$ represents the dimensionless frequency as defined in equation (\ref{ourODE}).} 
\label{table:1b}
\vspace*{1em}
\begin{tabular}{||c| c| c| c||} 
\hline
$\ell$ & 
$n$ & 
$\Omega$, $\mu=1.95$ &
$\Omega$, $\mu=2.25$\\ [0.5ex] 
\hline\hline
$2$ & $0$  & $0.3723-0.0776i$ & $0.3726-0.0861i$\\
    & $1$  & $0.3411-0.2316i$ & $0.3437-0.2611i$\\
    & $2$  & $0.2862-0.3945i$ & $0.2973-0.4462i$\\
    & $3$  & $0.2073-0.5902i$ & $0.2477-0.6492i$\\
$3$ & $0$  & $0.5979-0.0857i$ & $0.5985-0.0913i$\\
    & $1$  & $0.5730-0.2570i$ & $0.5786-0.2755i$\\
    & $2$  & $0.5252-0.4298i$ & $0.5417-0.4649i$\\
    & $3$  & $0.4569-0.6104i$ & $0.4923-0.6632i$\\
    & $4$  & $0.3693-0.8090i$ & $0.4347-0.8720i$
     \\[1ex] 
 \hline
 \end{tabular}
\end{table}

\begin{table}
\centering
\caption{This table presents the quasinormal frequencies for scalar perturbations ($s = 0$) of the extreme noncommutative geometry-inspired Schwarzschild black hole. The third and fourth columns showcase numerical values determined by \cite{Batic2019EPJC} using the WKB approximation and the inverted potential method (PT), respectively. The fifth column includes the results obtained via the spectral method employing $150$ polynomials with an accuracy of $200$ digits. Here, $\Omega$ represents the dimensionless frequency as defined in equation (\ref{ourODEe}).  The 'SM' abbreviation denotes the 'Spectral Method'.}
\label{table:2}
 \vspace*{1em}
 \begin{tabular}{||c| c| c| c| c||} 
 \hline
 $\ell$ & $n$ & $\Omega$ (WKB) \cite{Batic2019EPJC} & $\Omega$ (PT) \cite{Batic2019EPJC} & $\Omega$ (SM)\\ [0.5ex] 
 \hline\hline
$0$ & $0$ & $0.0395+0.1367i$ & $0.1003-0.1232i$ & $0.1027 - 0.0928i$\\ 
    & $1$ & $0.1524+0.4582i$ & $0.1003-0.3697i$ & $0.0299 - 0.3374i$\\
$1$ & $0$ & $0.2699-0.0744i$ & $0.2983-0.0997i$ & $0.2882 - 0.0895i$\\
    & $1$ & $0.1061-0.2684i$ & $0.2983-0.2991i$ & $0.2341 - 0.2777i$\\
    & $2$ & $0.1023+0.5721i$ & $0.2983-0.4985i$ & $0.1400 - 0.5021i$\\ 
    & $3$ & $0.3374+0.9027i$ & $0.2983-0.6979i$ & $0.0317 - 0.5720i$\\
$2$ & $0$ & $0.4756-0.0870i$ & $0.4881-0.0949i$ & $0.4802 - 0.0903i$\\
    & $1$ & $0.4136-0.2612i$ & $0.4881-0.2847i$ & $0.4447 - 0.2731i$\\
    & $2$ & $0.2782-0.4540i$ & $0.4881-0.4745i$ & $0.3760 - 0.4657i$\\
    & $3$ & $0.0918-0.6961i$ & $0.4881-0.6643i$ & $0.2806 - 0.6818i$\\ [0.5ex] 
 \hline
 \end{tabular}
\end{table}

\begin{table}
\centering
\caption{This table presents the quasinormal frequencies for electromagnetic perturbations ($s=1$) of the extreme noncommutative geometry-inspired Schwarzschild black hole via the spectral method employing $150$ polynomials with an accuracy of $200$ digits. Here, $\Omega$ represents the dimensionless frequency as defined in equation (\ref{ourODEe}).}
\label{table:3}
 \vspace*{1em}
 \begin{tabular}{||c| c| c||} 
 \hline
 $\ell$ & $n$ & $\Omega$\\ [0.5ex] 
 \hline\hline
$1$ & $0$ & $0.2465-0.0813i$\\ 
    & $1$ & $0.1904-0.2527i$ \\
    & $2$ & $0.0883-0.4627i$\\
$2$ & $0$ & $0.4549-0.0872i$ \\
    & $1$ & $0.4192-0.2640i$\\
    & $2$ & $0.3493-0.4503i$\\ 
    & $3$ & $0.2516-0.6601i$\\
 [0.5ex] 
 \hline
 \end{tabular}
\end{table}

\begin{table}
\centering
\caption{This table presents the quasinormal frequencies for vector-type gravitational perturbations ($s = 2$) of the extreme noncommutative geometry-inspired Schwarzschild black hole via the spectral method employing $150$ polynomials with an accuracy of $200$ digits. Here, $\Omega$ represents the dimensionless frequency as defined in equation (\ref{ourODEe}).}
\label{table:3b}
 \vspace*{1em}
 \begin{tabular}{||c| c| c||} 
 \hline
 $\ell$ & $n$ & $\Omega$\\ [0.5ex] 
 \hline\hline
$2$ & $0$ & $0.3723-0.0749i$\\ 
    & $1$ & $0.3394-0.2240i$\\
    & $2$ & $0.2811-0.3853i$\\
    & $3$ & $0.1975-0.5832i$\\
$3$ & $0$ & $0.5979-0.0839i$ \\
    & $1$ & $0.5711-0.2516i$ \\
    & $2$ & $0.5199-0.4214i$ \\ 
    & $3$ & $0.4471-0.6005i$ \\
    & $4$ & $0.3546-0.7997i$ \\
 [0.5ex] 
 \hline
 \end{tabular}
\end{table}

\section{Conclusions and outlook}

In this study, we have presented a comprehensive analysis of QNMs for noncommutative geometry-inspired Schwarzschild black holes, encompassing both non-extreme and extreme cases. Our investigation, employing the spectral method, covered scalar, electromagnetic, and gravitational perturbations. The significance of our findings is the remarkable stability exhibited by both non-extreme and extreme versions of these black holes across various perturbations. Contrary to previous claims in the literature, our results suggest that the noncommutative geometry-inspired Schwarzschild black holes are stable, a finding that is crucial for the understanding of their physical properties. This is particularly relevant in light of the existing debate over the stability of such black holes. Our study, therefore, provides a new perspective on the dynamics of these exotic objects and contributes to a more nuanced understanding of their behaviour under perturbations.

\subsection{Perspectives}

Our future endeavours will focus on expanding this study in several key directions. Firstly, we plan to investigate the QNMs in the context of massive perturbations. This extension will allow us to explore the influence of mass on the stability and dynamics of these black holes, offering further insights into their complex nature. Additionally, we aim to apply our methods to the study of QNMs of noncommutative geometry-inspired wormholes \cite{Garattini2009PLB, Nicolini2010CQG}. This will not only broaden our understanding of wormhole physics but also provide a comparative framework to evaluate the similarities and differences between black holes and wormholes within the context of noncommutative geometry. Another intriguing direction for future research involves the investigation of the so-called dirty black holes as proposed by \cite{Nicolini2010CQG}. These black holes, characterized by additional matter fields surrounding them, present a more realistic and complex scenario. Analyzing their QNMs will shed light on how external matter influences the stability and quasinormal spectra of black holes, further enriching our understanding of these objects. In summary, our research opens new avenues for exploring the intricate dynamics of noncommutative geometry-inspired black holes and wormholes. By extending our analysis to include massive perturbations and dirty black holes, we anticipate uncovering new aspects of these entities, contributing to the evolving landscape of gravitational and astrophysical research.

\section*{Code availability}

All analytical computations reported in this manuscript have been rechecked in the computer algebra system \textsc{Maple}. Two \textsc{Maple} sheets corresponding to the extreme and non-extreme cases can be found in supplementary materials \emph{and} in the repository below. The discretization of differential operators \eqref{L0nonee}--\eqref{L2nonee} using the Tchebyshev-type spectral method is equally performed in \textsc{Maple} computer algebra system. Finally, the numerical solution of the resulting quadratic eigenvalue problem \eqref{eq:eig} is performed in \textsc{Matlab} software using the \texttt{polyeig} function. All these materials are freely accessible at the following repository

\begin{itemize}
    \item \url{https://github.com/dutykh/noncommschwarzschild/}
\end{itemize}

\section*{Acknowledgements}

This publication is based upon work supported by the Khalifa University of Science and Technology under Award No. FSU$-2023-014$.

\bibliography{QNMS}

\begin{thebibliography}{43}
\expandafter\ifx\csname natexlab\endcsname\relax\def\natexlab#1{#1}\fi
\expandafter\ifx\csname bibnamefont\endcsname\relax
  \def\bibnamefont#1{#1}\fi
\expandafter\ifx\csname bibfnamefont\endcsname\relax
  \def\bibfnamefont#1{#1}\fi
\expandafter\ifx\csname citenamefont\endcsname\relax
  \def\citenamefont#1{#1}\fi
\expandafter\ifx\csname url\endcsname\relax
  \def\url#1{\texttt{#1}}\fi
\expandafter\ifx\csname urlprefix\endcsname\relax\def\urlprefix{URL }\fi
\providecommand{\bibinfo}[2]{#2}
\providecommand{\eprint}[2][]{\url{#2}}

\bibitem[{\citenamefont{Regge and Wheeler}(1957)}]{Regge1957PR}
\bibinfo{author}{\bibfnamefont{T.}~\bibnamefont{Regge}} \bibnamefont{and}
  \bibinfo{author}{\bibfnamefont{J.~A.} \bibnamefont{Wheeler}},
  \bibinfo{journal}{Phys. Rev.} \textbf{\bibinfo{volume}{108}},
  \bibinfo{pages}{1063} (\bibinfo{year}{1957}).

\bibitem[{\citenamefont{Hall}(2013)}]{Hall2013}
\bibinfo{author}{\bibfnamefont{B.~C.} \bibnamefont{Hall}},
  \emph{\bibinfo{title}{Quantum Theory for Mathematicians}}, vol.
  \bibinfo{volume}{267} of \emph{\bibinfo{series}{Graduate Texts in
  Mathematics}} (\bibinfo{publisher}{Springer: New York},
  \bibinfo{year}{2013}).

\bibitem[{\citenamefont{Ferrari and Mashhoon}(1984)}]{Ferrari1984PRD}
\bibinfo{author}{\bibfnamefont{V.}~\bibnamefont{Ferrari}} \bibnamefont{and}
  \bibinfo{author}{\bibfnamefont{B.}~\bibnamefont{Mashhoon}},
  \bibinfo{journal}{Phys. Rev. D} \textbf{\bibinfo{volume}{30}},
  \bibinfo{pages}{295} (\bibinfo{year}{1984}).

\bibitem[{\citenamefont{Leaver}(1985)}]{Leaver1985PRSLA}
\bibinfo{author}{\bibfnamefont{E.~W.} \bibnamefont{Leaver}},
  \bibinfo{journal}{Proc. R. Soc. Lond. A} \textbf{\bibinfo{volume}{402}},
  \bibinfo{pages}{285} (\bibinfo{year}{1985}).

\bibitem[{\citenamefont{Leaver}(1986)}]{Leaver1986PRD}
\bibinfo{author}{\bibfnamefont{E.~W.} \bibnamefont{Leaver}},
  \bibinfo{journal}{Phys. Rev. D} \textbf{\bibinfo{volume}{34}},
  \bibinfo{pages}{384} (\bibinfo{year}{1986}).

\bibitem[{\citenamefont{Nollert and Schmidt}(1992)}]{Nollert1992PRD}
\bibinfo{author}{\bibfnamefont{H.~P.} \bibnamefont{Nollert}} \bibnamefont{and}
  \bibinfo{author}{\bibfnamefont{B.~G.} \bibnamefont{Schmidt}},
  \bibinfo{journal}{Phys.Rev. D} \textbf{\bibinfo{volume}{45}},
  \bibinfo{pages}{2617} (\bibinfo{year}{1992}).

\bibitem[{\citenamefont{Ciftci et~al.}(2003)\citenamefont{Ciftci, Hall, and
  Saad}}]{Ciftci2003JPA}
\bibinfo{author}{\bibfnamefont{H.}~\bibnamefont{Ciftci}},
  \bibinfo{author}{\bibfnamefont{R.~L.} \bibnamefont{Hall}}, \bibnamefont{and}
  \bibinfo{author}{\bibfnamefont{N.}~\bibnamefont{Saad}}, \bibinfo{journal}{J.
  Phys. A: Math. Gen.} \textbf{\bibinfo{volume}{36}}, \bibinfo{pages}{11807}
  (\bibinfo{year}{2003}).

\bibitem[{\citenamefont{Cho et~al.}(2010)\citenamefont{Cho, Cornell, Doukas,
  and Naylor}}]{Cho2010CQG}
\bibinfo{author}{\bibfnamefont{H.~T.} \bibnamefont{Cho}},
  \bibinfo{author}{\bibfnamefont{A.}~\bibnamefont{Cornell}},
  \bibinfo{author}{\bibfnamefont{J.}~\bibnamefont{Doukas}}, \bibnamefont{and}
  \bibinfo{author}{\bibfnamefont{W.}~\bibnamefont{Naylor}},
  \bibinfo{journal}{Class. Quantum Grav.} \textbf{\bibinfo{volume}{27}},
  \bibinfo{pages}{155004} (\bibinfo{year}{2010}).

\bibitem[{\citenamefont{Cho et~al.}(2012)\citenamefont{Cho, Cornell, Doukas,
  Huang, and Naylor}}]{Cho2012AMP}
\bibinfo{author}{\bibfnamefont{H.~T.} \bibnamefont{Cho}},
  \bibinfo{author}{\bibfnamefont{A.}~\bibnamefont{Cornell}},
  \bibinfo{author}{\bibfnamefont{J.}~\bibnamefont{Doukas}},
  \bibinfo{author}{\bibfnamefont{T.-R.} \bibnamefont{Huang}}, \bibnamefont{and}
  \bibinfo{author}{\bibfnamefont{W.}~\bibnamefont{Naylor}},
  \bibinfo{journal}{Adv. Math. Phys.} \textbf{\bibinfo{volume}{2012}},
  \bibinfo{pages}{281705} (\bibinfo{year}{2012}).

\bibitem[{\citenamefont{Batic et~al.}(2018)\citenamefont{Batic, Nowakowski, and
  Redway}}]{Batic2018PRD}
\bibinfo{author}{\bibfnamefont{D.}~\bibnamefont{Batic}},
  \bibinfo{author}{\bibfnamefont{M.}~\bibnamefont{Nowakowski}},
  \bibnamefont{and} \bibinfo{author}{\bibfnamefont{K.}~\bibnamefont{Redway}},
  \bibinfo{journal}{Phys. Rev. D} \textbf{\bibinfo{volume}{98}},
  \bibinfo{pages}{024017} (\bibinfo{year}{2018}).

\bibitem[{\citenamefont{Macedo}(2019)}]{Panosso2019PRD}
\bibinfo{author}{\bibfnamefont{R.~P.} \bibnamefont{Macedo}},
  \bibinfo{journal}{Phys. Rev. D} \textbf{\bibinfo{volume}{99}},
  \bibinfo{pages}{088501} (\bibinfo{year}{2019}).

\bibitem[{\citenamefont{Batic et~al.}(2019{\natexlab{a}})\citenamefont{Batic,
  Nowakowski, and Redway}}]{Batic2019PRDa}
\bibinfo{author}{\bibfnamefont{D.}~\bibnamefont{Batic}},
  \bibinfo{author}{\bibfnamefont{M.}~\bibnamefont{Nowakowski}},
  \bibnamefont{and} \bibinfo{author}{\bibfnamefont{K.}~\bibnamefont{Redway}},
  \bibinfo{journal}{Phys. Rev. D} \textbf{\bibinfo{volume}{99}},
  \bibinfo{pages}{088502} (\bibinfo{year}{2019}{\natexlab{a}}).

\bibitem[{\citenamefont{Ciftci et~al.}(2005)\citenamefont{Ciftci, Hall, and
  Saad}}]{Ciftci2005PLA}
\bibinfo{author}{\bibfnamefont{H.}~\bibnamefont{Ciftci}},
  \bibinfo{author}{\bibfnamefont{R.~L.} \bibnamefont{Hall}}, \bibnamefont{and}
  \bibinfo{author}{\bibfnamefont{N.}~\bibnamefont{Saad}},
  \bibinfo{journal}{Phys. Lett. A} \textbf{\bibinfo{volume}{340}},
  \bibinfo{pages}{388} (\bibinfo{year}{2005}).

\bibitem[{\citenamefont{Mamani et~al.}(2022)\citenamefont{Mamani, Masa,
  Sanches, and Zanchin}}]{Mamani2022EPJC}
\bibinfo{author}{\bibfnamefont{L.~A.~H.} \bibnamefont{Mamani}},
  \bibinfo{author}{\bibfnamefont{A.~D.~D.} \bibnamefont{Masa}},
  \bibinfo{author}{\bibfnamefont{L.~T.} \bibnamefont{Sanches}},
  \bibnamefont{and} \bibinfo{author}{\bibfnamefont{V.~T.}
  \bibnamefont{Zanchin}}, \bibinfo{journal}{Eur. Phys. J. C}
  \textbf{\bibinfo{volume}{82}}, \bibinfo{pages}{897} (\bibinfo{year}{2022}).

\bibitem[{\citenamefont{Iyer}(1987)}]{IYER1987PRD}
\bibinfo{author}{\bibfnamefont{H.~S.} \bibnamefont{Iyer}},
  \bibinfo{journal}{Phys. Rev. D} \textbf{\bibinfo{volume}{35}},
  \bibinfo{pages}{3632} (\bibinfo{year}{1987}).

\bibitem[{\citenamefont{Daghigh and Green}(2012)}]{Daghigh2012PRD}
\bibinfo{author}{\bibfnamefont{R.~G.} \bibnamefont{Daghigh}} \bibnamefont{and}
  \bibinfo{author}{\bibfnamefont{M.~D.} \bibnamefont{Green}},
  \bibinfo{journal}{Phys. Rev. D} \textbf{\bibinfo{volume}{85}},
  \bibinfo{pages}{127501} (\bibinfo{year}{2012}).

\bibitem[{\citenamefont{Batic and Nowakowski}(2023)}]{Batic2023MMAS}
\bibinfo{author}{\bibfnamefont{D.}~\bibnamefont{Batic}} \bibnamefont{and}
  \bibinfo{author}{\bibfnamefont{M.}~\bibnamefont{Nowakowski}},
  \bibinfo{journal}{Math. Methods Appl. Sci.} \textbf{\bibinfo{volume}{46}},
  \bibinfo{pages}{10833} (\bibinfo{year}{2023}).

\bibitem[{\citenamefont{Nicolini et~al.}(2006)\citenamefont{Nicolini,
  Smailagic, and Spallucci}}]{Nicolini2006PLB}
\bibinfo{author}{\bibfnamefont{P.}~\bibnamefont{Nicolini}},
  \bibinfo{author}{\bibfnamefont{A.}~\bibnamefont{Smailagic}},
  \bibnamefont{and}
  \bibinfo{author}{\bibfnamefont{E.}~\bibnamefont{Spallucci}},
  \bibinfo{journal}{Phys. Lett. B} \textbf{\bibinfo{volume}{632}},
  \bibinfo{pages}{547} (\bibinfo{year}{2006}).

\bibitem[{\citenamefont{Bachelot and
  Motet-Bachelot}(1993)}]{Bachelot1993AIHPPT}
\bibinfo{author}{\bibfnamefont{A.}~\bibnamefont{Bachelot}} \bibnamefont{and}
  \bibinfo{author}{\bibfnamefont{A.}~\bibnamefont{Motet-Bachelot}},
  \bibinfo{journal}{Ann. Inst. H. Poincare Phys. Theor.}
  \textbf{\bibinfo{volume}{59}}, \bibinfo{pages}{3} (\bibinfo{year}{1993}).

\bibitem[{\citenamefont{Gundlach et~al.}(1994)\citenamefont{Gundlach, Price,
  and Pullin}}]{Gundlach1994PRD}
\bibinfo{author}{\bibfnamefont{C.}~\bibnamefont{Gundlach}},
  \bibinfo{author}{\bibfnamefont{R.~H.} \bibnamefont{Price}}, \bibnamefont{and}
  \bibinfo{author}{\bibfnamefont{J.}~\bibnamefont{Pullin}},
  \bibinfo{journal}{Phys. Rev. D} \textbf{\bibinfo{volume}{49}},
  \bibinfo{pages}{883} (\bibinfo{year}{1994}).

\bibitem[{\citenamefont{Andersson}(1995)}]{Andersson1995PRD}
\bibinfo{author}{\bibfnamefont{N.}~\bibnamefont{Andersson}},
  \bibinfo{journal}{Phys.Rev. D} \textbf{\bibinfo{volume}{51}},
  \bibinfo{pages}{353} (\bibinfo{year}{1995}).

\bibitem[{\citenamefont{Andersson}(1997)}]{Andersson1997PRD}
\bibinfo{author}{\bibfnamefont{N.}~\bibnamefont{Andersson}},
  \bibinfo{journal}{Phys.Rev. D} \textbf{\bibinfo{volume}{55}},
  \bibinfo{pages}{468} (\bibinfo{year}{1997}).

\bibitem[{\citenamefont{Konoplya}(2003)}]{Konoplya2003ATMP}
\bibinfo{author}{\bibfnamefont{R.~A.} \bibnamefont{Konoplya}},
  \bibinfo{journal}{Adv. Theor. Math. Phys.} \textbf{\bibinfo{volume}{6}},
  \bibinfo{pages}{1135} (\bibinfo{year}{2003}).

\bibitem[{\citenamefont{Motl and Neitzke}(2003)}]{Motl2003ATMP}
\bibinfo{author}{\bibfnamefont{L.}~\bibnamefont{Motl}} \bibnamefont{and}
  \bibinfo{author}{\bibfnamefont{A.}~\bibnamefont{Neitzke}},
  \bibinfo{journal}{Adv. Theor. Math. Phys.} \textbf{\bibinfo{volume}{7}},
  \bibinfo{pages}{307} (\bibinfo{year}{2003}).

\bibitem[{\citenamefont{Natário and Schiappa}(2004)}]{Natario2004ATMP}
\bibinfo{author}{\bibfnamefont{J.}~\bibnamefont{Natário}} \bibnamefont{and}
  \bibinfo{author}{\bibfnamefont{R.}~\bibnamefont{Schiappa}},
  \bibinfo{journal}{Adv. Theor. Math. Phys.} \textbf{\bibinfo{volume}{8}},
  \bibinfo{pages}{1001} (\bibinfo{year}{2004}).

\bibitem[{\citenamefont{Chakrabarti and Gupta}(2006)}]{Chakrabarti2006IJMPA}
\bibinfo{author}{\bibfnamefont{S.~K.} \bibnamefont{Chakrabarti}}
  \bibnamefont{and} \bibinfo{author}{\bibfnamefont{K.~S.} \bibnamefont{Gupta}},
  \bibinfo{journal}{Int. J .Mod. Phys. A} \textbf{\bibinfo{volume}{21}},
  \bibinfo{pages}{3565} (\bibinfo{year}{2006}).

\bibitem[{\citenamefont{Giri}(2007)}]{Giri2007IJMPA}
\bibinfo{author}{\bibfnamefont{P.~R.} \bibnamefont{Giri}},
  \bibinfo{journal}{Int. J. Mod. Phys. A} \textbf{\bibinfo{volume}{22}},
  \bibinfo{pages}{2047} (\bibinfo{year}{2007}).

\bibitem[{\citenamefont{Batic et~al.}(2019{\natexlab{b}})\citenamefont{Batic,
  Kelkar, Nowakowski, and Redway}}]{Batic2019EPJC}
\bibinfo{author}{\bibfnamefont{D.}~\bibnamefont{Batic}},
  \bibinfo{author}{\bibfnamefont{N.~G.} \bibnamefont{Kelkar}},
  \bibinfo{author}{\bibfnamefont{M.}~\bibnamefont{Nowakowski}},
  \bibnamefont{and} \bibinfo{author}{\bibfnamefont{K.}~\bibnamefont{Redway}},
  \bibinfo{journal}{Eur. Phys. J. C} \textbf{\bibinfo{volume}{79}},
  \bibinfo{pages}{581} (\bibinfo{year}{2019}{\natexlab{b}}).

\bibitem[{\citenamefont{Liang}(2018{\natexlab{a}})}]{Liang2018CPLa}
\bibinfo{author}{\bibfnamefont{J.}~\bibnamefont{Liang}},
  \bibinfo{journal}{Chinese Phys. Lett.} \textbf{\bibinfo{volume}{35}},
  \bibinfo{pages}{010401} (\bibinfo{year}{2018}{\natexlab{a}}).

\bibitem[{\citenamefont{Liang}(2018{\natexlab{b}})}]{Liang2018CPLb}
\bibinfo{author}{\bibfnamefont{J.}~\bibnamefont{Liang}},
  \bibinfo{journal}{Chinese Phys. Lett.} \textbf{\bibinfo{volume}{35}},
  \bibinfo{pages}{050401} (\bibinfo{year}{2018}{\natexlab{b}}).

\bibitem[{\citenamefont{Batic and Nicolini}(2010)}]{Batic2010PLB}
\bibinfo{author}{\bibfnamefont{D.}~\bibnamefont{Batic}} \bibnamefont{and}
  \bibinfo{author}{\bibfnamefont{P.}~\bibnamefont{Nicolini}},
  \bibinfo{journal}{Phys.Lett.B} \textbf{\bibinfo{volume}{692}},
  \bibinfo{pages}{32} (\bibinfo{year}{2010}).

\bibitem[{\citenamefont{Brown and Mann}(2011)}]{Brown2011PLB}
\bibinfo{author}{\bibfnamefont{E.}~\bibnamefont{Brown}} \bibnamefont{and}
  \bibinfo{author}{\bibfnamefont{R.}~\bibnamefont{Mann}},
  \bibinfo{journal}{Phys. Lett. B} \textbf{\bibinfo{volume}{694}},
  \bibinfo{pages}{440} (\bibinfo{year}{2011}).

\bibitem[{\citenamefont{Abramowitz and Stegun}(1972)}]{Abramowitz1972}
\bibinfo{author}{\bibfnamefont{M.}~\bibnamefont{Abramowitz}} \bibnamefont{and}
  \bibinfo{author}{\bibfnamefont{I.~A.} \bibnamefont{Stegun}},
  \emph{\bibinfo{title}{Handbook of Mathematical Functions}}
  (\bibinfo{publisher}{Dover: New York}, \bibinfo{year}{1972}).

\bibitem[{\citenamefont{Ince}(1956)}]{Ince1956}
\bibinfo{author}{\bibfnamefont{E.~L.} \bibnamefont{Ince}},
  \emph{\bibinfo{title}{Ordinary Differential Equations}}
  (\bibinfo{publisher}{Dover: New York}, \bibinfo{year}{1956}).

\bibitem[{\citenamefont{Bender and Orszag}(1999)}]{Bender1999}
\bibinfo{author}{\bibfnamefont{C.~M.} \bibnamefont{Bender}} \bibnamefont{and}
  \bibinfo{author}{\bibfnamefont{S.~A.} \bibnamefont{Orszag}},
  \emph{\bibinfo{title}{Advanced Mathematical Methods for Scientists and
  Engineers I: Asymptotic Methods and Perturbation Theory}}
  (\bibinfo{publisher}{Springer: New York}, \bibinfo{year}{1999}).

\bibitem[{\citenamefont{Olver}(1994)}]{Olver1994MAA}
\bibinfo{author}{\bibfnamefont{F.~W.~J.} \bibnamefont{Olver}},
  \bibinfo{journal}{Methods Appl. Anal.} \textbf{\bibinfo{volume}{1}},
  \bibinfo{pages}{1} (\bibinfo{year}{1994}).

\bibitem[{\citenamefont{Trefethen}(2000)}]{Trefethen2000}
\bibinfo{author}{\bibfnamefont{L.~N.} \bibnamefont{Trefethen}},
  \emph{\bibinfo{title}{Spectral methods in MatLab}}
  (\bibinfo{publisher}{Society for Industrial and Applied Mathematics,
  Philadelphia, PA, USA}, \bibinfo{year}{2000}),
  \urlprefix\url{http://web.comlab.ox.ac.uk/oucl/work/nick.trefethen/spectral.html}.

\bibitem[{\citenamefont{Boyd}(2000)}]{Boyd2000}
\bibinfo{author}{\bibfnamefont{J.~P.} \bibnamefont{Boyd}},
  \emph{\bibinfo{title}{Chebyshev and Fourier Spectral Methods}}
  (\bibinfo{publisher}{Dover Publications, New York}, \bibinfo{year}{2000}),
  \bibinfo{edition}{2nd} ed.

\bibitem[{\citenamefont{Fox and Parker}(1968)}]{Fox1968}
\bibinfo{author}{\bibfnamefont{L.}~\bibnamefont{Fox}} \bibnamefont{and}
  \bibinfo{author}{\bibfnamefont{I.~B.} \bibnamefont{Parker}},
  \emph{\bibinfo{title}{Chebyshev Polynomials in Numerical Analysis}}
  (\bibinfo{publisher}{Oxford University Press}, \bibinfo{year}{1968}).

\bibitem[{\citenamefont{Tisseur and Meerbergen}(2001)}]{Tisseur2001}
\bibinfo{author}{\bibfnamefont{F.}~\bibnamefont{Tisseur}} \bibnamefont{and}
  \bibinfo{author}{\bibfnamefont{K.}~\bibnamefont{Meerbergen}},
  \bibinfo{journal}{SIAM Review} \textbf{\bibinfo{volume}{43}},
  \bibinfo{pages}{235} (\bibinfo{year}{2001}), ISSN \bibinfo{issn}{0036-1445}.

\bibitem[{\citenamefont{Holodoborodko}(2023)}]{mct2015}
\bibinfo{author}{\bibfnamefont{P.}~\bibnamefont{Holodoborodko}},
  \emph{\bibinfo{title}{{Multiprecision Computing Toolbox for MATLAB
  5.2.5.15444}}} (\bibinfo{publisher}{Advanpix LLC.},
  \bibinfo{address}{Yokohama, Japan}, \bibinfo{year}{2023}).

\bibitem[{\citenamefont{Garattini and Lobo}(2009)}]{Garattini2009PLB}
\bibinfo{author}{\bibfnamefont{R.}~\bibnamefont{Garattini}} \bibnamefont{and}
  \bibinfo{author}{\bibfnamefont{F.~S.~N.} \bibnamefont{Lobo}},
  \bibinfo{journal}{Phys. Lett. B} \textbf{\bibinfo{volume}{671}},
  \bibinfo{pages}{146} (\bibinfo{year}{2009}).

\bibitem[{\citenamefont{Nicolini and Spallucci}(2010)}]{Nicolini2010CQG}
\bibinfo{author}{\bibfnamefont{P.}~\bibnamefont{Nicolini}} \bibnamefont{and}
  \bibinfo{author}{\bibfnamefont{E.}~\bibnamefont{Spallucci}},
  \bibinfo{journal}{Class. Quant. Grav.} \textbf{\bibinfo{volume}{27}},
  \bibinfo{pages}{015010} (\bibinfo{year}{2010}).

\end{thebibliography}

\end{document}